\documentclass[]{mn2e}
\usepackage{amssymb}

\newif\ifAMStwofonts
\def\ARAA{ARA\&A}

\def\ee #1 {\times 10^{#1}}
\def\ut #1 #2 { \, \rmn{#1}^{#2}}
\def\u #1 { \, \rmn{#1}}

\def\micron {\, \mu \hbox{m}}
\def\half{{\textstyle \frac{1}{2}}}
\def\thalf{{\textstyle{ 3\over 2}}}
\let\grad=\nabla
\def\cross{\bmath{\times}}
\def\curl #1 {\grad \cross #1}
\def\div #1 {\grad \cdot #1}

\def\e{\bmath{e}}
\def\v{\bmath{v}}

\def\B{\bmath{B}}
\def\E{\bmath{E}}            % E
\def\Bh{\bmath{\hat{B}}}
\def\fh{\bmath{\hat{\phi}}}  % phi
\def\rh{\bmath{\hat{r}}}     % r
\def\vr{v_{r}}
\def\vf{v_{\phi}}            % v_phi
\def\vk{v_{K}}               % v_K

\def\Epa{\bmath{E'_\parallel}}  % E'_||
\def\Epe{\bmath{E'_\perp}}  % E'_perp
\def\J{\bmath{J}}

\def\dv{\bmath{\delta\v}}
\def\dE{\bmath{\delta\E}}
\def\dB{\bmath{\delta\B}}
\def\dJ{\bmath{\delta\J}}

\newcommand{\delt} [1] {\frac{\partial #1}{\partial t}}

\newcommand{\delz} [1] {\frac{\partial #1}{\partial z}}
\input epsf

\title{Magnetorotational instability in stratified, weakly ionised 
accretion discs}
\author[R. Salmeron and M. Wardle]
       {Raquel Salmeron$ ^1 $ \& Mark Wardle$ ^2 $ \\
$ ^1 $School of Physics, University of Sydney, NSW 2006, Australia \\
$ ^2 $Physics Department, Macquarie University, NSW 2109, Australia}

\date{2003 March 20}
\pagerange{\pageref{firstpage}--\pageref{lastpage}}
\pubyear{2003}
\begin{document}
\maketitle
\label{firstpage}
\begin{abstract}
We present a linear analysis of the vertical structure and growth of the
magnetorotational instability in stratified, weakly ionised accretion discs,
such as protostellar and quiescent dwarf novae systems.
The method includes the effects of the magnetic coupling, the conductivity
regime of the fluid and the strength of the magnetic field, which is initially
vertical. The conductivity is treated as a tensor and assumed constant with
height.

We obtained solutions for the structure and growth rate of global unstable 
modes for different conductivity regimes, strengths of the initial 
magnetic field and coupling between ionised and neutral components of the 
fluid.
The envelopes of short-wavelength perturbations 
are determined by the action of competing local growth rates at 
different heights, driven by the vertical stratification of the disc. 
Ambipolar diffusion perturbations 
peak consistently higher above the midplane than modes including Hall 
conductivity. For weak coupling, perturbations including the Hall effect 
grow faster and act over a more extended cross-section of the disc than 
those obtained using the ambipolar diffusion approximation.
  
Finally, we derived an approximate criterion for when Hall diffusion 
determines the 
growth of the magnetorotational instability. This is satisfied over a wide 
range of radii in protostellar discs, reducing the extent of the magnetic 
`dead zone'. Even if the magnetic coupling is weak, significant accretion 
may occur close to the midplane, rather than in the surface regions of 
weakly-ionised discs.

\end{abstract}

\section{Introduction}
The collapse of protostellar cores lead to the  development of a
central mass or protostar, surrounded by a disc of material, which is
accreted towards the center. During this process angular momentum
is transferred to a small percentage of disc material at large radii,
enabling the collapse of most of the disc towards the central star
(e.g. Weintraub, Sandell \& Duncan 1989; Adams, Emerson \& Fuller 1990;
Beckwith et al. 1990).
The evolution of this `disc accretion' phase is dependent upon the rate of
angular momentum transport in the disc (e.g. Adams \& Lin 1993).

A variety of mechanisms have been invoked to explain this
transport. As the molecular viscosity of accretion discs is too low to
explain observed accretion rates (Pringle 1981), some form of turbulent
viscosity must be present. The origin and characteristics of this turbulence
remains an important problem in star formation theories. Convective turbulence
 has been considered as an option (Lin \& Papaloizou 1980), but further
studies suggest that this mechanism may transport angular momentum
towards the central star instead of away from it (Cabot \& Pollack 1992;
Ryu \& Goodman 1992; Stone \& Balbus 1996).  The gravitational field of a
companion star may trigger hydrodynamic waves which can transport
angular momentum (Vishniac \& Diamond 1989; Rozyczka \& Spruit 1993), but
as a significant fraction of stars do not belong to binary systems, this
mechanism is not general enough to explain accretion processes in all stars.

Balbus and Hawley have pointed out that the nature of this anomalous viscosity
can be hydromagnetic (Balbus \& Hawley 1991; Hawley \& Balbus
1991; Stone et al. 1996). This `magnetorotational' instability (MRI) had been
described initially by Velikov (1959) and Chandrasekhar (1961) through their
analysis of magnetised Couette flows. It drives turbulent motions which
transport angular momentum radially outwards,
as fluid elements exchange angular momentum non-locally by means of the
distortion of the magnetic field lines that connect them.

Under ideal MHD conditions, MRI perturbations grow in discs that are
differentially rotating, with the angular velocity increasing outwards.
Axisymmetric modes need a magnetic field with a weak, poloidal component. In
this context, `weak' means that the magnetic
energy density of the field is less than the thermal energy density. These
perturbations have a characteristic length scale
$\lambda \sim v_A/ \Omega$, where $v_A$ is the Alfv\'en speed and $\Omega$ is
the keplerian angular frequency in the disc, and
a maximum growth rate  $\sim q\Omega/2$, with
$q \equiv 1.5$ for keplerian discs (Balbus \& Hawley 1992a).
This growth rate does not depend on the strength or direction of the
magnetic field as long as a poloidal component is present.
Non-axisymmetric perturbations are most unstable under the influence of a
poloidal field, but also grow at a reduced rate if the field is purely
toroidal (Balbus \& Hawley 1992b). These perturbations are of interest for the
analysis of field amplification mechanisms, as dynamo amplification can not
occur through axisymmetric perturbations (Moffatt 1978). With no
strong dissipation processes, no other conditions are required. Because of
its robustness and the general conditions under which it develops,
the magnetorotational instability is a promising source of turbulent
viscosity in accretion discs.

Ideal MHD conditions are a good approximation to model astrophysical systems
where the ionisation fraction of the gas is high enough to ensure neutral and
ionised components of the fluid are well coupled. Active dwarf novae (except
possibly in the outer regions) and black hole accretion discs are
examples of such systems.
However, in dense, cool environments such as those of protostellar
discs, it is doubtful that magnetic coupling is significant over the
entire radial and vertical dimensions of the discs (Gammie 1996; Wardle 1997).
Similar conditions are thought to apply in quiescent and the outer regions
of hot-state dwarf novae discs (Gammie \& Menou 1998; Menou 2000; Stone et al.
2000). In these cases, low conductivity significantly affects the growth and
structure of MRI
perturbations. Different approximations have been adopted to account for the
departure
from ideal MHD in low conductivity astrophysical discs (see section
\ref{subsec:governing}).

Most models of the MRI in low conductivity discs have
used the ambipolar diffusion (Blaes \& Balbus 1994; MacLow et al. 1995;
Hawley \& Stone 1998) or resistive (Jin 1996; Balbus \& Hawley 1998; Sano,
Inutsuka \& Miyama 1998) limits.
Recently, it has been recognised the importance of the Hall conductivity terms
in addition to resistivity for the analysis of low conductivity discs (Wardle
1999, hereafter W99, Balbus \& Terquem 2001 and Sano \& Stone
2002a,b; 2003).

The huge
variation of fluid variables over the vertical and radial extension of
astrophysical discs is a further complication. Vertical stratification is
particularly relevant, as these objects
are generally thin and changes in the plane of the disc are much more gradual
than those in the direction perpendicular to it. Previous models of the MRI 
have not included density stratification and Hall conductivity simultaneously.
It is expected that solutions will be strongly modified when both factors are 
present. This motivates the present study. 

This paper examines the structure and linear growth of the magnetorotational
instability in vertically stratified, non self-gravitating accretion discs.
We assume the disc is isothermal and
geometrically thin, so variations in the fluid variables in the radial
direction can be ignored. The initial magnetic field is vertical and
the analysis is restricted to perturbations with wavevector perpendicular to
the plane of the disc ($k=k_z$). These are the most unstable perturbations
with the adopted field geometry, as
magnetic pressure strongly supresses displacements with $k_r \neq 0$ (Balbus
\& Hawley 1991, Sano \& Miyama 1999). The
conductivity of the gas is treated as a tensor and assumed constant with
height in this initial study, although the formulation is
also valid for a $z$-dependent conductivity.
This makes the present method a powerful tool
for the analysis of more realistic discs (see section \ref{sec:discussion} - 
Discussion).

Section \ref{sec:formulation}
presents the governing equations for a weakly ionised, magnetised
disc in near-keplerian motion around the central star and details the
adopted disc model.
Section \ref{sec:linearisation} summarises the linearisation of the
equations and presents the final linear system in dimensionless form.
It also describes the three parameters that control the dynamics of the
fluid. Section \ref{sec:boundary conditions} discusses the boundary
conditions used to integrate the equations from the midplane to
the surface of the disc and the integration method. Section
\ref{sec:results} presents the test cases used to characterise the
conductivity regimes relevant for this work and compares our results with a
previous local analysis. It also details key findings on the dependency
of the structure and growth rate of the perturbations with the conductivity
regime, the strength of the magnetic field and its coupling with the
neutral gas. These results are discussed in section
\ref{sec:discussion}. By way of example, this section also
calculates the structure and growth rate of the MRI under fluid
conditions where different conductivity regimes are dominant at different
heights above the midplane.
This reflects (qualitatively) the conditions expected to be found in
real discs. Finally, the methodology and key findings of this paper are
summarised in section \ref{sec:summary}.

\section{Formulation}
\label{sec:formulation}

\subsection{Governing Equations}
\label{subsec:governing}

The equations of non-ideal MHD are written about a local keplerian frame
corotating with the disc at the angular frequency $\Omega$ associated with
the particular radius of interest. Consequently, the velocity of the fluid
can be expressed as a departure from exact keplerian motion
$\bmath{v} = \bmath{V} - \v_K$, where $\bmath{V}$ is the velocity in the
standard laboratory coordinate system $(r, \phi, z)$ anchored at the central
mass $M$, and $\v_K = \sqrt {GM/r} \fh$ is the keplerian
velocity at the radius $r$. Similarly, if $\partial / \partial t$ is
the time derivative in the local keplerian frame, then the time derivative in
the laboratory frame can be expressed as
$\partial / \partial t + \Omega \partial / \partial \phi$. We also assume that
the fluid is weakly ionised, meaning that the abundances of charged species
are so low that their inertia and thermal pressure, as well as the effect of 
ionisation and recombination processes in the neutral gas, are negligible. 
These
assumptions effectively restrict the range of frequencies that can be studied
with this formulation to be smaller than the collision frequency of any of the
charged species with the neutrals. Accordingly, separate equations of motion
for the charged species are not required and their effect on the neutrals is
contained in a conductivity tensor (see section \ref{subsec:conductivity}).

The governing equations are the continuity equation,

\begin{equation}
\delt{\rho} + \div(\rho \v) = 0 \,,
	\label{eq:continuity}
\end{equation}

\noindent
the equation of motion,

\begin{eqnarray}
\lefteqn{\delt{\v} + (\v \cdot \grad)\v -2\Omega \vf \rh + \half\Omega \vr\fh
-\frac{\vk^2}{r}\rh + \frac{c_s^2}{\rho}\grad\rho + {} }
	\nonumber\\
& & {}+\grad \Phi - \frac{\J\cross\B}{c\rho} = 0\,,
	\label{eq:momentum}
\end{eqnarray}

\noindent
and the induction equation,

\begin{equation}
\delt{\B} = \curl (\v \cross \B) - c \curl \E' -\thalf \Omega \B_r \fh \,.
	\label{eq:induction}
\end{equation}

\noindent

In the equation of motion (\ref{eq:momentum}), $\Phi$ is the gravitational
potential due to the central gas, given by

\begin{equation}
\Phi = -\frac{GM}{(r^2 + z^2)^{\half}} \,,
          \label{eq:gravpotential}
\end{equation}

\noindent
and $\v_K^2/r$ is the centripetal term generated by exact keplerian
motion. At the disc midplane this term balances the radial component
of the gravitational potential. The terms $2\Omega \vf\rh$ and $\half \Omega
\vr\fh$ are the coriolis terms associated with the use of a local keplerian
frame, $c_s = \sqrt {P/\rho}$ is the isothermal sound speed,
$\Omega = \v_K/r$ is the keplerian frequency and $c$ is the speed
of light. Other symbols have their usual meanings.

In the induction equation (\ref{eq:induction}), the term $c \curl \E'$
contains the effects of non-ideal MHD. $\E'$ is the electric field in the
frame comoving with the neutrals and the term $\thalf \Omega \B_r \fh$
accounts for the generation of toroidal field from the radial component due to
the differential rotation of the disc.

Additionally, the magnetic field must satisfy the constraint:

\begin{equation}
\div \B = 0 \,,
\label{eq:divB}
\end{equation}

\noindent
and the current density must satisfy Ampere's law,

\begin{equation}
\J = \frac{c}{4\pi}\grad\cross \B %\,,
\label{eq:j_curlB}
\end{equation}

\noindent
and Ohm's law,

\begin{equation}
	\J = \bmath{\sigma}\cdot \E' \,.
	\label{eq:ohm}
\end{equation}

Note that the conductivity, which depends on the abundance and drifts
of the charged species through the neutral gas is treated as a tensor
$\bmath{\sigma}$, as detailed in the following section. This formulation is
compared to the drift velocity approach in section \ref{subsec:multifluid}.

\subsection{The conductivity tensor $\bmath{\sigma}$}
\label{subsec:conductivity}

The electric conductivity is a tensor whenever the gyrofrequency of the
charged
carriers is larger than the frequency of momentum exchange by collisions with
the neutrals, or $|\beta|\gg 1$ (Cowling 1957, Norman \& Hayvaerts 1985,
Nakano \& Umebayashi 1986). On the contrary, when collisions with the neutrals
are dominant the conductivity is a scalar, the ordinary ohmic resistivity. To
obtain expressions for the components of this tensor, we begin by writing down
the equations of motion of the ionised species.
As inertia and
thermal pressure are neglected, the motion of the charged particles is given
by the balance of the Lorentz force and the drag force from
collisions with the neutrals,

\begin{equation}
Z_j e\left(\E' + {\v_j \over c} \cross \B\right) - \gamma_j m_j \v_j
= 0 \,,
	\label{eq:drift}
\end{equation}

\noindent
where each charged species $j$ is characterised by its number density $n_j$,
particle mass $m_j$, charge $Z_j e$ and drift velocity $\v_j$. In the above
equation,

\begin{equation}
	\gamma_j = \frac{<\sigma v>_j}{m_j+m} \,,
	\label{eq:gamma}
\end{equation}

\noindent
where $m$ is the mean mass of the neutral particles and $<\sigma v>_j$ is the
rate coefficient of momentum exchange by collisions with the neutrals. We
will also make use of the Hall parameter,

\begin{equation}
\beta_j= {Z_jeB \over m_j c} \, {1 \over \gamma_j \rho}\, ,
\label{eq:Hall_parameter}
\end{equation}

\noindent
given by the ratio of the gyrofrequency and the collision frequency of
charged species
$j$ with the neutrals. It represents the relative importance of the Lorentz
and drag terms in equation (\ref{eq:drift}).

Following the treatment of Wardle \& Ng (1999) and W99 we use the following
expression for Ohm's law,

\begin{equation}
	\J = \bmath{\sigma}\cdot \E' = \sigma_{\parallel} \Epa +
	\sigma_1 \Bh \cross \Epe + \sigma_2 \Epe  \,,
	\label{eq:J-E}
\end{equation}

\noindent
obtained by inverting
(\ref{eq:drift}) to express $\v_j$ as a function of $\E'$ and
$\B$ and then using $\J = e\sum_j n_j Z_j \v_j$ together with the charge
neutrality assumption $\sum_j n_j Z_j = 0$.
In equation (\ref{eq:J-E}), $\Epa$ and $\Epe$ are the
components of the electric field $\E'$, parallel and perpendicular to the
magnetic field, respectively. The
components of the conductivity tensor $\bmath{\sigma}$ are the conductivity
parallel to the magnetic field,

\begin{equation}
	\sigma_{\parallel} = \frac{ec}{B}\sum_{j} n_j Z_j \beta_j \,,
	\label{eq:sigma0}
\end{equation}

\noindent
the Hall conductivity,

\begin{equation}
	\sigma_1 = \frac{ec}{B}\sum_{j}\frac{n_j Z_j}{1+\beta_j^2}\,,
	\label{eq:sigma1}
\end{equation}

\noindent
and the Pedersen conductivity,

\begin{equation}
	\sigma_2 = \frac{ec}{B}\sum_{j}\frac{n_j Z_j \beta_j}{1+\beta_j^2} \,.
	\label{eq:sigma2}
\end{equation}

The relative values of the components of the conductivity tensor
differentiate  three conductivity regimes:

\begin{enumerate}
\item The \emph{ambipolar diffusion} regime occurs when
$\sigma_\parallel \gg \sigma_2 \gg |\sigma_1|$, or $|\beta| \gg 1$ for most
charged
species. This implies that most charged particles are strongly tied to the
magnetic field by electromagnetic stresses. This regime is dominant at
relatively low densities, where the magnetic field is frozen into the ionised
component of the fluid and drifts with it through the neutrals.
The linear behaviour of the MRI
in this regime has been analysed by Blaes \& Balbus (1994) and
the non-linear growth by MacLow et al. (1995) and Hawley \& Stone (1998).

\item The \emph{resistive (Ohmic)} regime is obtained when most charged
species are linked to the neutrals via collisions. This occurs when
$\sigma_\parallel \approx \sigma_2 \gg |\sigma_1|$, implying $|\beta| \ll 1$.
This regime is predominant closest to the midplane, where the high
density makes the collision frequency of the charged particles with the
neutrals high enough to prevent the former from drifting. This case has been
studied under a linear approximation by Jin (1996), Balbus \& Hawley (1998),
Papaloizou \& Terquem (1997), Sano \& Miyama (1999) and Sano et. al. (2000).
The non-linear regime has been studied by Sano et al. (1998);
Fleming, Stone \& Hawley (2000) and  Stone \& Fleming, (2003). This last work
includes a $z$-dependent resistivity.

\item  Finally, the \emph{Hall} regime occurs when charged particles of one
sign are tied to the magnetic field while those of the other sign follow the
neutrals. In this case $|\sigma_1| \sim \sigma_2 < \sigma_{\parallel}$ and
$|\beta| \approx 1$. It is important at intermediate densities,
between the ones associated with the ambipolar and Ohmic difussion regimes.
Recent studies have explored the MRI with Hall
effects in the linear (W99, Balbus \& Terquem 2001) and non-linear
regimes (Sano \& Stone 2002a,b; 2003).
\end{enumerate}

\subsection{Comparison with the multifluid approach}
	\label{subsec:multifluid}

Another commonly used form of the induction equation is obtained
by assuming that ions and electrons are the main charge carriers and
drift through the neutrals
 (e.g Balbus \& Terquem 2001, Sano \& Stone 2002). Using
this approach, the induction equation is

\begin{eqnarray}
\lefteqn{\delt{\B} = \curl \left (\v_e \cross \B\right) }
	\nonumber\\
& & {} = \curl \left [\v \cross \B  - \frac{4 \pi \eta \J}{c} -
\frac{\J \cross \B}{e n_e} +
\frac{(\J \cross \B) \cross \B}{c \gamma_i \rho \rho_i} \right]
	\nonumber\\
& & {}
	\label{eq:induction_drift}
\end{eqnarray}

\noindent
where $\v_e$ is the electron drift speed, $\eta=c^2/4 \pi \sigma$ is the 
resistivity and subscripts $e$ and $i$
refer to electrons and ions, respectively. The four terms in the right
hand side of equation (\ref{eq:induction_drift}) are, from left to right, the
inductive, resistive, Hall and ambipolar diffusion terms.
We now express equation (\ref{eq:induction}) in terms of $\J$, $\B$ and the
components of the conductivity tensor in order to show that, in
the appropriate limit, it corresponds to equation
(\ref{eq:induction_drift}), as expected.
We begin by inverting equation
(\ref{eq:J-E}) to find an expression for $\E'$,

\begin{equation}
	\E' = \frac{\J}{\sigma_\parallel} +
	      \frac{\sigma_1}{\sigma_\perp^2} \frac{\J\cross\B}{B} -
	\left(\frac{\sigma_2}{\sigma_\perp^2}-
	\frac{1}{\sigma_\parallel}\right)
	\frac{(\J\cross\B)\cross\B}{B^2}
	\label{eq:E-J}
\end{equation}

\noindent
where $\sigma_{\perp} = \sqrt{\sigma_1^2 + \sigma_2^2}$ is the total
conductivity perpendicular to the magnetic field.
Assuming that the only charged species are ions and
electrons with Hall parameters $\beta_i$ and $\beta_e$ ($<0$),
respectively, and that charge neutrality is satisfied
($n_i = n_e$), we obtain the following expressions for the components of the
conductivity tensor:

\begin{equation}
	\sigma_\parallel = \frac{cen_e}{B} (\beta_i - \beta_e) \,,
	\label{eq:sigp}
\end{equation}

\begin{equation}
	\sigma_1 = \frac{cen_e}{B} \frac{(\beta_i + \beta_e)(\beta_e -
	\beta_i)}{(1+\beta_e^2)(1+\beta_i^2)} \,,\textrm{\ and}
	\label{eq:sig1}
\end{equation}

\begin{equation}
	\sigma_2 = \frac{cen_e}{B} \frac{(1-\beta_i\beta_e)(\beta_i -
	\beta_e)}{(1+\beta_e^2)(1+\beta_i^2)} \,.
	\label{eq:sig2}
\end{equation}

\noindent
From (\ref{eq:sig1}) and (\ref{eq:sig2}), we find:

\begin{equation}
	\sigma_\perp = \frac{cen_e}{B} \frac{(\beta_i -
	\beta_e)}{[(1+\beta_e^2)(1+\beta_i^2)]^{1/2}} \,.
	\label{eq:sigperp}
\end{equation}

\noindent
Substituting these expressions into (\ref{eq:E-J}) gives

\begin{eqnarray}
\lefteqn{\E' = \frac{\beta_e}{\beta_i-\beta_e}\,\frac{\J}{\sigma} +
	      \frac{\beta_e+\beta_i}{\beta_e-\beta_i} \,
\frac{\J\cross\B}{c e n_e} -
	       \frac{\beta_e}{\beta_e -
\beta_i}\,\frac{(\J\cross\B)\cross\B}{c^2
\gamma_i\rho\rho_i}}
	\nonumber\\
& & {}
	\label{eq:E-J2}
\end{eqnarray}

\noindent
where $\sigma$ is the electrical conductivity due to electrons.
Finally, simplifying (\ref{eq:E-J2}) by using $|\beta_e| \gg \beta_i$ (as is
the case) and
substituting the resulting $\E'$ in
equation (\ref{eq:induction}) (without the coriolis term), yields the
standard result, shown in (\ref{eq:induction_drift}).

Each of the last three terms in the right hand side of equation
(\ref{eq:induction_drift}) dominate when the fluid is in a particular
conductivity regime (section \ref{subsec:conductivity}). These limits can
also be recovered with
the appropriate assumptions, through equations (\ref{eq:E-J}) and
(\ref{eq:E-J2}).
To get the resistive regime, for example, we substitute $\sigma_1=0$ and
$\sigma_{\parallel} = \sigma_2$ into equation (\ref{eq:E-J}). In this
limit the conductivity is a scalar (the resistive
approximation), so $\E' = \J/\sigma$ and the induction equation reduces to the
familiar form,

\begin{equation}
\delt{\B} = \curl \left [\v \cross \B - \frac{c^2}{4 \pi \sigma}
\curl \B \right]\,.
%-\thalf \Omega \B_r \fh \,.
	\label{eq:induction_3}
\end{equation}

On the other hand, to model the Hall limit, we regard the charged species to
be either `ions',
which are strongly tied to the neutrals through collisions ($\beta_i \ll 1$),
or `electrons', for which the only important forces are electromagnetic
stresses ($|\beta_e| \gg 1$) (see also discussion in W99). In this limit the
Hall
conductivity is

\begin{equation}
|\sigma_1| = \frac{c e n_e}{B}
	\label{eq:sigma_hall_1}
\end{equation}

\noindent
while the Pedersen conductivity is

\begin{equation}
\sigma_2 = |\sigma_1| \left (\beta_i - \frac{1}{\beta_e} \right)
\ll \sigma_1 \,.
	\label{eq:sigma_hall_2}
\end{equation}

As the ions are effectively locked with the neutrals, the
current density will be given by the drift of the electrons through the
neutral gas. Collisions are unimportant in their equation of motion, so they
drift perpendicular to
the plane of the electric and magnetic fields, in order to annul the Lorentz
force acting upon them. In this limit $(\J \cross \B) \cross \B = -\J$ and
\begin{equation}
\E'= \frac{\J \cross \B}{c e n_e} \,,
	\label{eq:E-Hall}
\end{equation}
consistent with the Hall term in equation (\ref{eq:induction_drift}).

Finally, the ambipolar diffusion approximation is recovered by assuming
$\sigma_{\parallel} \gg \sigma_2$, $\sigma_1=0$ and
$|\beta_e| \gg \beta_i \gg 1$.
In this limit, the Pedersen conductivity is given by

\begin{equation}
\sigma_2 =  \frac{ce}{B}\frac{n_i}{\beta_i} \,, \textrm{\ and}
	\label{eq:ad_sigma_2}
\end{equation}

\begin{equation}
\E' = - \frac{(\J \cross \B) \cross \B}{c^2 \gamma_i \rho \rho_i} \,,
	\label{eq:ad_term}
\end{equation}

\noindent
which originates the ambipolar diffusion term of (\ref{eq:induction_drift}).

Although the multifluid drift and conductivity tensor formulations are
ultimately equivalent, which one is more
convenient depends on the problem at hand. In particular, the presence of
dust grains tends to make the treatment of different species
especially complex.
In protostellar discs, dust grains can be the
more abundant charged species over extended regions.
For example, assuming $0.1 \micron$ grains, negatively charged grains
dominate whenever $n_H \gtrsim 10^{11}$ $\textrm{cm}^{-3}$, while positive
charged grains are the most abundant ions for
$n_H \gtrsim 10^{14}$ $\textrm{cm}^{-3}$ (Wardle \& Ng 1999). Having separate
equations of motion for different
charged species would generally involve
dividing the grain size distribution of interest into an appropriate number of
discrete intervals and explicitly treating each
one. Unless the number of such intervals is small it is easy to see that this
method can become very cumbersome. In these circumstances, incorporating the
contribution of each charged species into a conductivity tensor
can be a valuable approach.

\subsection{Disc Model}
	\label{subsec:model}

Our model incorporates the vertical structure of the disc, but neglects fluid
variations in the radial direction. This is appropriate as astrophysical
accretion discs are generally thin and changes in the radial direction occur
in a much bigger length scale than those in the vertical direction.
Including the vertical structure means that perturbations of spatial
dimensions
comparable to the scale height of the disc, which are associated with a
strong magnetic field ($v_A \sim c_s$), or low conductivity, can be explored.

The balance between the vertical component of the central
gravitational force and the pressure gradient within the disc determines
its equilibrium structure. The vertical density distribution
in hydrostatic equilibrium is given by

\begin{equation}
\frac{\rho (r,z)}{\rho_{o}(r)}= \exp \left(-\frac{z^2}{2 H^2(r)}\right) \,.
	\label{eq:rhoinitial}
\end{equation}

\noindent
In the above equation, $\rho_{o}(r)$ is the gas density at the midplane and
$H(r) = c_s/\Omega$ is the scaleheight of the disc.

A self-consistent treatment of this problem, would involve adopting a
particular dependency of $\rho_o$ and $H$ with $r$ using a suitable model,
such as the minimum solar nebula (Hayashi, Nakazawa \& Nakagawa 1985) and
calculating $\rho(r,z)$ by means of (\ref{eq:rhoinitial}).
This density, together with the adopted strength of the magnetic field
$\bmath{B}$
and the values of the conductivity tensor $\bmath{\sigma}$ as a function of
height would be used to evaluate the
parameters that govern the fluid evolution (see section
\ref{subsec:parameters}) and solve the fluid equations.

The realistic evaluation of the conductivity tensor is a
complex undertaking, as it depends critically on the abundances of charged
species (ions, electrons and charged dust grains) which, in turn, are a
function of the ionisation balance in the disc. This balance is given by the
equilibrium between ionisation processes by cosmic rays, radioactive
elements and
X-rays from the central star (e.g. Hayashi 1981; Glassgold, Najita \& Igea
1997,
Igea \& Glassgold 1999; Fromang, Terquem \& Balbus 2002)
and recombinations taking place both in the gas phase and in grain surfaces
(e.g. Nishi, Nakano \& Umebayashi 1991). In low conductivity discs, the
level of ionisation is insufficient to produce good coupling
between the magnetic field and the neutral component of the fluid over their
entire vertical structure. In protostellar discs, for example, it is expected
that in the region
outside $\sim 0.1$ AU from the central star, the coupling will be significant
only in the surface layers, where X-rays and cosmic rays can penetrate and
ionise the gas (Gammie 1996; Wardle 1997). In these environments, the
$z$-dependent atenuation of the
ionisation rate typical of interstellar space $10^{-17} \textrm{H}^{-1}$
$\textrm{s}^{-1}$ has to be taken into account.

On the other hand, the contribution of dust grains to recombination processes
is particularly complex, because they generally have a distribution of
sizes and corresponding collision cross sections (Mathis, Rumpl \& Nordsieck
1977; Umebayashi \& Nakano 1990). Moreover, the dynamics of grain particles
depends on the activity of
the disc. In quiescent environments, they tend to settle towards the
midplane and begin to agglomerate into bigger structures that could
eventually become planets (e.g. Weidenschilling \& Cuzzi 1993). This removes
grains from regions at relatively high $z$ and causes the ionisation level to
increase by reducing recombination processes taking place in their surfaces.
Simulations of the evolution of dust grain distributions
in dense cores show that they grow icy mantles and coagulate efficiently
(Ossenkopf 1993). By this process, the smallest particles grow quickly while
the upper grain size limit is only slightly changed (Ossenkopf 1993). As a
result, the surface area of dust grains can be significantly modified by
grain evolution, which ultimately affects the ionisation balance in the disc.

In the present work, a simpler treatment has been adopted. In order to
study the MRI under different conductivity regimes, the values of the
components of
the conductivity tensor have been selected so that they satisfy the conditions
outlined in section \ref{subsec:conductivity} for each regime. For simplicity,
these values are assumed to be constant, although the formulation
allows them to be a function of height.  We found that the parameters that
control the evolution of the fluid are the ratio of the components of the
conductivity tensor perpendicular to the magnetic field
($\sigma_1/ \sigma_2$), the strength of the magnetic field and its degree of
coupling with the neutral component of the fluid (see section
\ref{subsec:parameters}). The midplane values of these parameters have been
selected in order to simulate the fluid conditions we were interested in
modelling. This approach will be discussed in section \ref{subsec:test} where
the chosen test cases are detailed.

With this approach we are able to study the dependency of the growth rate and
structure (characterised by the height of maximum amplitude and the
wavenumber of unstable modes) of the instability with the parameters
of the fluid in a stratified disc. This is
relevant, as the region where linear perturbations peak
is also expected to be the region where non-linear perturbations grow fastest,
until turbulence finally sets in and causes all wavenumbers to interact so
that, eventually, the longest wavelengths carry the greatest angular momentum
transport (e.g. Hawley \& Balbus 1995).
%Finally, through this study we are able to determine
%the region of parameter space where unstable modes grow and the dependency of
%the growth rate with the parameters of the fluid.

\section{LINEARISATION}
\label{sec:linearisation}

We linearised the system of equations (\ref{eq:continuity})
- (\ref{eq:induction}), (\ref{eq:j_curlB}) and (\ref{eq:J-E}) about an initial
steady state where
$\J = \v = \E' = 0$ and $\B = B \hat{z}$.
In the initial state both $\E'$ and $\J$ vanish, so the changes in the
conductivity tensor due
to the perturbations do not appear in the linearised equations. As a result,
it is not necessary to explore how the perturbations affect the
conductivity and only the values in the initial steady state are required.

\subsection{Linearised Equations}
	\label{subsec:linearised}

We assume the wavevector of the perturbations is perpendicular to the plane
of the disc ($k = k_z$). Perturbations with a vertical wavenumber, initiated
from a vertically aligned equilibrium magnetic field, exhibit the fastest
growth rate for
a given set of parameters because magnetic pressure strongly
suppresses displacements with $k_r \neq 0$ (Balbus \& Hawley 1991; Sano \&
Miyama 1999). Taking perturbations of the form
$\bmath{q} = \bmath{q}_{0} + \bmath{\delta q}(z) \e^{i\omega t}$ about the
initial state, linearising
and neglecting terms of order $H/r$ or smaller, we find that the equations
decouple into two subsystems. One of them corresponds to sound waves
propagating in the vertical direction and the other describes MHD
perturbations in the plane of the disc, with vanishing $z$ component.

With these simplifying assumptions, the final linear system of equations that
describes the MHD perturbations within the disc is,

\begin{equation}
i\omega\rho \dv_r - 2 \rho \Omega \dv_{\phi} - \frac{B_0}{c}\dJ_{\phi} = 0 \,,
	\label{eq:motion_r_final}
\end{equation}

\begin{equation}
i\omega\rho \dv_{\phi} + \half\rho \Omega \dv_r + \frac{B_0}{c} \dJ_r = 0 \,,
	\label{eq:motion_phi_final}
\end{equation}

\begin{equation}
i\omega \dB_r - c\delz{\dE_{\phi}} = 0 \,,
	\label{eq:induc_r_final}
\end{equation}

\begin{equation}
i\omega \dB_{\phi} + c \delz{\dE_r} + \thalf\Omega \dB_r = 0 \,,
	\label{eq:induc_phi_final}
\end{equation}

\begin{equation}
\dJ_r = - \frac{c}{4 \pi}\delz{\dB_{\phi}} \,,
	\label{ampere_r_final}
\end{equation}

\begin{equation}
\dJ_{\phi} = \frac{c}{4\pi}\delz{\dB_r} \,,
	\label{eq:ampere_phi_final}
\end{equation}

\vspace{0.14cm}

\begin{equation}
\dJ_r = \sigma_2 \dE_r' - \sigma_1\dE_{\phi}' \,,
	\label{eq:ohm_r_final}
\end{equation}

\vspace{0.14cm}

\begin{equation}
\dJ_{\phi} = \sigma_1 \dE_r' + \sigma_2\dE_{\phi}' \,,
	\label{eq:phm_phi_final}
\end{equation}

\noindent
where $\dE_{\phi}$ and $\dE_r$ are the perturbations of the electric field in
the laboratory frame, given by

\begin{equation}
\dE_{\phi} = \dE_{\phi}' + \frac{B_0}{c}\dv_r \,, \textrm{\ and}
	\label{eq:E'phi_v}
\end{equation}

\begin{equation}
\dE_r = \dE'_r - \frac{B_0}{c}\dv_{\phi} \,.
	\label{eq:E'r_v}
\end{equation}

We note that $\sigma_{\parallel}$, the component of the conductivity tensor
parallel to the magnetic field, does not appear in the final linearised
equations. Because the ambipolar diffusion and resistive conductivity regimes
are differentiated by the value of $\sigma_{\parallel}$ (section
\ref{subsec:conductivity}), this means that during the linear stage of the
MRI, under the adopted approximations, these regimes are identical.

\subsection{Equations in dimensionless form}
	\label{subsec:dimensionless}

Equations (\ref{eq:motion_r_final}) to (\ref{eq:E'r_v}) can be expressed in
dimensionless form, normalising the variables as follows:

\begin{displaymath}
z^{*} = \frac{z}{H} \qquad
\rho^{*} = \frac{\rho(r,z)}{\rho_0(r)} \qquad
\dB^* = \frac{\dB}{\B_0} \qquad
\end{displaymath}

\vspace{0.3cm}

\begin{displaymath}
\dv^* = \frac{\dv}{c_s} \qquad
\dE^* = \frac{c \dE}{c_s \bmath{B_0}} \qquad
\dE'^* = \frac{c \dE'}{c_s \bmath{B_0}} \qquad
\end{displaymath}

\vspace{0.3cm}

\begin{displaymath}
\dJ^* = \frac{c \dJ}{c_s \bmath{B_0}\sigma_{\perp_0}} \qquad
\bmath{\sigma}^{*} = \frac{\bmath{\sigma}}{\sigma_{\perp_0}} \qquad
\bmath{\sigma_{\perp}^{*}} = \frac{\sigma_{\perp}}{\sigma_{\perp_0}}
\end{displaymath}

Subscript `o' is used to
denote variables at the midplane of the disc. Effecting these changes and
dropping the asterisks to keep the notation simple, we finally express the
dimensionless system of equations in matrix form as:

\begin{equation}
\frac{d}{dz}\left( \begin{array}{c}
B_r \\
\\
B_{\phi} \\
\\
E^{}_r \\
\\
E^{}_{\phi}
\end{array} \right)
 = \left( \begin{array}{cccc}
0 & 0 & C_1 A_1 & C_1 A_2 \\
\\
0 & 0 & -C_1 A_2 & C_1 A_3 \\
\\
-\thalf & -\nu & 0 & 0\\
\\
\nu & 0 & 0 & 0
\end{array} \right) \left( \begin{array}{c}
B_r\\
\\
B_{\phi}\\
\\
E_r\\
\\
E_{\phi}
\end{array} \right)
	\label{eq:matrix}
\end{equation}

\vspace{10pt}

\begin{equation}
\dJ = \sigma_{\perp}C_2 \left( \begin{array}{cc}
A2 & -A3\\
\\
A1 & A2
\end{array} \right) \dE
	\label{eq:matrix_JE}
\end{equation}

\vspace{10pt}

\begin{equation}
\dv = \frac{\chi_o}{\rho} \frac{1}{1 + \nu^2}
\left( \begin{array}{cc}
-2 & \nu\\
\\
-\nu & -\half
\end{array} \right) \dJ
	\label{eq:matrix_vJ}
\end{equation}

\vspace{10pt}

\begin{equation}
\dE' = \frac{1}{{\sigma_{\perp}}^2}\left( \begin{array}{cc}
\sigma_2^{} & \sigma_1^{}\\
\\
-\sigma_1^{} & -\sigma_2^{}
\end{array} \right)\dJ
	\label{eq:matrix_EJ}
\end{equation}

\noindent
where %$\sigma_1^{'} = \sigma_1/\sigma_{\bot}$,
%$\sigma_2^{'} = \sigma_2/\sigma_{\bot}$, $\sigma_{\perp0}$ is the value of
%$\sigma_{\perp}$ at the midplane,

\vspace{8pt}

\begin{equation}
\nu =\frac{i\omega}{\Omega} \,,
	\label{eq:nu}
\end{equation}

\begin{equation}
C_1 = \chi_o \sigma_{\perp} \left(\frac{v_A}{c_s}\right)^{-2} C_2 \,,
	\label{eq:C1}
\end{equation}

\begin{equation}
C_2 = \left[1+\frac{\chi_o \sigma_{\perp}}{\rho}\frac{1}{1+\nu^2}
\left(\frac{5}{2}\frac{\sigma_1}{\sigma_{\perp}} +
2 \nu \frac{\sigma_2}{\sigma_{\perp}} + \frac{\chi_o \sigma_{\perp}}{\rho}
\right)\right]^{-1} \,,
	\label{eq:C0}
\end{equation}

\begin{equation}
A_1 = \frac{\sigma_1}{\sigma_{\perp}} + 2\frac{\chi_o \sigma_{\perp}}{\rho}
\frac{1}{1 + \nu^2} \,,
	\label{eq:A1}
\end{equation}

\begin{equation}
A_2 = \frac{\sigma_2}{\sigma_{\perp}} + \nu \frac{\chi_o \sigma_{\perp}}{\rho}
\frac{1}{1 + \nu^2} \,, \textrm{\ and}
	\label{eq:A2}
\end{equation}

\begin{equation}
A_3 = \frac{\sigma_1}{\sigma_{\perp}} + \frac{1}{2}\frac{\chi_o \sigma_{\perp}}
{\rho}\frac{1}{1 + \nu^2} \,.
	\label{eq:A3}
\end{equation}

\noindent
In the above expressions,

\begin{equation}
v_A = \frac{B_o}{\sqrt {4 \pi \rho_o}}
	\label{eq:alfven}
\end{equation}

\noindent
is the Alfv\'en speed at the midplane of the disc, and

\begin{equation}
\chi_o = \frac{\omega_{co}}{\Omega} = \frac{1}{\Omega}\frac{B_o^2
\sigma_{\perp o}}{\rho_o c^2}
	\label{eq:chi}
\end{equation}

\noindent
is a parameter that characterises the midplane coupling between the magnetic
field and the disc (see section \ref{subsec:parameters}). To understand the
information contained in this parameter it is useful to recall that the
effect of finite conductivity is different for perturbations of different
wavelengths. Finite conductivity is important when the term $c \curl \E'$ \ in
the induction equation (\ref{eq:induction}) is of order $\curl (\v
\cross \B)$.\  Adopting a length scale $L \sim 1/k \sim v_A/ \omega$ and
$\sigma \sim \sigma_{\perp}$, it is found that these two terms are comparable
when  $kv_A \sim (c^2/4 \pi) k^2/ \sigma_{\perp}$, in other
words, non-ideal effects will strongly modify wavemodes at, or above, the
critical frequency,

\begin{equation}
\omega_c = \frac{B^2 \sigma_{\perp}}{\rho c^2} \,.
	\label{eq:omega_c}
\end{equation}

It can be shown (W99) that in the limit $|\beta_j| \rightarrow \infty$,
$\omega_c$ reduces to $\sum_j \gamma_j \rho_j$, the collision frequency of
the neutrals with any of the charged species. Generally, $\omega_c$ is
smaller than this value and much smaller than $\gamma_j \rho$, the
collision frequency of charged species $j$ with the neutrals. In dense
clouds, $\omega_c \sim \gamma_G \rho_G$, which is the smallest $\gamma_j
\rho_j$, the
collision frequency of neutrals with grains. As $\rho_G \sim 0.01 \rho$, the
treatment of this paper, restricted to $\omega < \gamma_j \rho$ by neglecting
the inertia of the charged species, remains valid for $\omega \sim
100\omega_c$.
For perturbations with lower frequencies (longer wavelength) than $\omega_c$,
ideal MHD (the flux-freezing approximation) is valid.

\subsection{Parameters}
\label{subsec:parameters}

As these equations reveal, three important parameters control the evolution
of the fluid:

\begin{enumerate}
\item $v_A/c_s$, the ratio of the Alfv\'en speed to the isothermal sound
speed of the gas at the midplane. It is a measure of the strength of the
magnetic field. In ideal MHD unstable modes grow when the magnetic field is
subthermal ($v_A/c_s < 1$). When $v_A \sim c_s$ the minimum
wavelength of the instability is of the order of the scaleheight of the disc
and the growth rate decreases rapidly.

\item $\chi_o$, a parameter that characterises the strength of the coupling
between the magnetic field and the disc at the midplane (see equation
\ref{eq:chi}). It is given by the ratio of the critical frequency above
which flux-freezing
conditions break down and the dynamical frequency of the disc at the
midplane. If $\chi_o = \omega_{co}/\Omega < 1$ the
disc is poorly coupled to the disc at the
frequencies of interest for dynamical analysis.
As the growth rate of the most unstable modes are of order $\Omega$ in ideal
MHD conditions, these are also the interesting frequencies for the study of
this instability.

\item $\sigma_1/\sigma_2$, the ratio of the conductivity terms
perpendicular to the magnetic field. It is an indication of the
conductivity regime of the fluid, as discussed in section
\ref{subsec:conductivity}.
\end{enumerate}

Note that the density of the disc decreases with $z$, so the local values of
$\chi$ and $v_A/c_s$ increase with height. The parameters of the model
are defined as the corresponding values at the midplane.

It is common practice to characterise the magnetic coupling of a weakly
ionised
fluid by its electron density $n_e$. Before finishing this section, we discuss
how $\chi$ relates to this fluid parameter. We begin by writing the
magnetic Reynolds number as (e.g. Balbus \& Terquem 2001),

\begin{equation}
	\mathrm{Re_M} = \frac{v_A H}{\eta} =
	%\frac{4\pi\sigma_\parallel}{c^2}\, \frac{c_s}{\Omega} =
	\frac{B^2\sigma_\parallel}{\rho c^2} \,
	\frac{\sqrt{4\pi\rho}}{B} \, \frac{c_s}{\Omega} =
	\frac{\sigma_\parallel}{\sigma_\perp} \, \frac{\chi}{v_A / c_s} \,,
	\label{eq:Rem}
\end{equation}

\noindent
where we have used $\eta = c^2/4 \pi \sigma_{\parallel}$.
If ions and electrons are the only charged species, then

\begin{equation}
	\chi =  \frac{v_A}{c_s} \,\frac{\mathrm{Re_M}}{(1+\beta_e^2)^{1/2}
(1+\beta_i^2)^{1/2}} \,.
	\label{eq:chi1}
\end{equation}

In the resistive regime, $|\beta| \ll 1$ for both charged species and
(\ref{eq:chi1}) shows that the criterion for non-Hall MRI
perturbations to grow (W99), $\chi > v_A/c_s$, is equivalent to
$\mathrm{Re_M > 1}$.

We can now obtain an expression for $\chi$ in terms of the electron fraction
$x_e = n_e/n_H$ at the midplane,

\begin{eqnarray}
	\chi = \frac{1}{\Omega}\frac{B^2 \sigma_{\perp}}{\rho c^2} & = &
	\frac{en_e B}{c\Omega\rho}\,\frac{\beta_i -
	\beta_e}{(1+\beta_e^2)^{1/2}(1+\beta_i^2)^{1/2}}
	\nonumber\\
        & \approx &
	%\frac{en_e B}{c\Omega\rho}\,\frac{|\beta_e|}{(1+\beta_e^2)^{1/2}} =
	\frac{ n_e<\sigma v> m_e}{(m_e+m_n) \Omega}
	\,\frac{\beta_e^2}{(1+\beta_e^2)^{1/2}}\,,
	\label{eq:chi-ne}
\end{eqnarray}

\noindent
where we used expression (\ref{eq:sigperp}) for $\sigma_\perp$ and assumed
$\beta_i \ll |\beta_e|$.  In the above equation

\begin{equation}
	<\sigma v> \approx 1\ee -15 \left(\frac{128 kT}{9\pi
	m_e}\right)^{1/2} \ut cm -2
	\label{eq:sigv}
\end{equation}

\noindent
is the momentum-transfer rate coefficient for electron-neutral
scattering.  Note that the dependence of $\chi$ on magnetic field strength now
enters only through the electron Hall parameter $\beta_e$.
Following Fromang, Terquem \& Balbus (2003)
we assume that grains have settled out and the
electron (and ion) number density is determined by the recombination of
metal ions given by,

\begin{equation}
	n_e \approx \left(\frac{\zeta n_\mathrm{H}}{\alpha}\right)^{1/2} \,,
	\label{eq:ne}
\end{equation}

\noindent
where $\alpha \approx 3\ee -11 T^{-1/2} \ut cm 3 \ut s -1 $ is the
radiative recombination rate for metal ions.  The ionisation rate
$\zeta$ is assumed to be due to cosmic rays at a rate
$10^{-17} \exp (-\Sigma / 96 \u g \ut cm -2 )\ut s -1 \ut H -1 $ where
$\Sigma$ is the disc surface density.  This dominates x-ray
ionisation for the column densities we shall consider here. Results are shown
in Table \ref{table:ionisation} for a nominal 1 solar mass star and $B=10\u
mG $.

\begin{table*}
\caption{Comparison of the magnetic coupling parameter $\chi_o$ and the
ionisation fraction $x_e$ at the midplane for different radial positions
$r_o$, with $B = 10$ \ mG and assuming grains have settled out (Fromang,
Terquem \& Balbus 2002). Also shown are the assumed temperature $T_o$ and
calculated values of $n_H$, $v_A/c_s$, $\zeta$, $\beta_e$ and $\beta_i$.}

\begin{tabular}{|c|r|l|l|l|l|l|l|l|}
\hline
$r_o$ (AU) & $T_o$ (K) & $n_H$ ($\textrm{cm}^{-3}$) & $v_A/c_s$ &
 $\zeta$ ($\textrm{s}^{-1} \textrm{H}^{-1}$) &
 $|\beta_e|$ & $\beta_i$ & $x_e$ & $\chi_o$ \\
\hline
$1$ & $280$ & $6 \ee 14 $ & $0.00076$ &  $5.76 \ee -22 $ & $0.035$ &
           	$7.68 \ee -5 $ & $7.32 \ee -13 $ & $8.9 \ee -6 $   \\
$5$ & $130$ & $7 \ee 12 $ &$0.010$ & $4.81 \ee -18 $ &
	$4.44$ & $6.58 \ee -3 $ & $5.11 \ee -10 $ & $1.9$ \\
$10$ & $90$ & $1 \ee 12 $ & $0.033$ & $9.76 \ee -18 $ &
	$37.31$ & $0.046$ & $1.76 \ee -9 $ & $19$  \\
\hline
\end{tabular}
	\label{table:ionisation}
\end{table*}

For this strength of the magnetic field, $|\beta_e| \ll 1$ at 1 AU and $\chi$
scales as $B^2$ (up to about 200 mG) at this radius. On the other hand,
$\beta_e$ is greater
than $1$ at $5$ and $10$ AU, so $\chi$ will scale linearly with $B$ (see
equation \ref{eq:chi-ne}).  In particular for $B=100$ mG at $1$
AU, $\chi \approx 0.00088$, consistent with the detailed calculations in
Wardle (2003).

\section{Boundary Conditions}
\label{sec:boundary conditions}

To solve equations (\ref{eq:matrix}) to (\ref{eq:matrix_EJ}) it is necessary
first to integrate the system of ordinary differential equations (ODE) in
(\ref{eq:matrix}).
This problem can be treated as a two-point boundary value problem for coupled
ODE. Five boundary conditions must be formulated, prescribed either at the
midplane or at the surface of the disc.

\begin{description}
\item \textbf{At the midplane}: A set of boundary conditions can be arrived
at by asuming fluid variables have either `odd' or `even' symmetry about
the midplane. `Odd' symmetry means the variable is an odd function of $z$
and vanishes at $z=0$. Conversely, when `even' symmetry is applied, the
variable is assumed to be an even function of $z$ and its gradient is zero at
the midplane. In this paper we applied the odd - even symmetry criteria to the
perturbations in the magnetic field, $\dB(z) = \pm \dB(-z)$, where the upper
(lower) sign corresponds to even (odd) symmetry conditions. This contrasts
with Lovelace, Wang \& Sulkanen (1987), who applied the symmetry criteria to
the flux function $\Psi(r,z) = rA_{\phi}$ with $A_{\phi}$ the toroidal
component of the vector potential and obtained $\dB_{r,\phi} (r,z) = \mp
\dB_{r, \phi} (r, -z)$ as their symmetry criteria.
The symmetry of a particular fluid variable is assigned arbitrarily, subject
to the constraint that fluid equations are satisfied. This means that two sets
of boundary conditions are equally valid, obtained by reversing the assumed
symmetry of the fluid variables.  Perturbations obtained with a particular
set of boundary conditions are displaced a quarter of a wavelength from those
found
with the other one. The growth rates of these solutions lie at intermediate
points of the curve $\nu$ vs $k$ obtained from the local analysis (W99), as
expected. Evidently, no
generality is lost by focusing in one of these two possible sets of
solutions. We compare the growth rate versus number of nodes of perturbations
obtained with `odd' and `even' symmetry in section
\ref{subsec:comparison} (Comparison with local analysis). For the rest of the
analysis presented in this paper we chose to assign odd symmetry to
$\delta B_r$ and $\delta B_{\phi}$, so they vanish at $z=0$. This gives us
two boundary conditions at the midplane.
As the equations are linear, their overall scaling is arbitrary,
so a
third boundary condition can be obtained by setting one of the
fluid variables to any convenient value. To that effect, we assigned
a value of $1$
to $\delta E'_r$. Summarising, three boundary conditions are applied at the
midplane:

\begin{displaymath}
\dB_r = \dB_{\phi} = 0, \textrm{\ and}
\end{displaymath}

\begin{displaymath}
\dE'_r = 1 \,.
\end{displaymath}

\item \textbf{At the surface}: At sufficiently high $z$ above the midplane,
ideal MHD conditions hold. This assumption is
appropriate in this case because the local coupling parameter $\chi$ is
inversely proportional to the
density and so it is stronger at higher $z$ regions where the density is
smaller. When $\chi>10$ the growth rate and characteristic wavenumber of
unstable modes differ little from the ideal limit (W99), so even
though for simplicity we have assumed the conductivity tensor to be spatially
constant, we can assume that flux-freezing conditions hold at the surface
and the local dispersion relation is $k v_A = \Omega$
(Balbus \& Hawley 1991). As
the Alfv\'en speed increases with $\rho^{1/2}$, the wavelengths
of magnetic field perturbations increase with $z$ and given the dependency of
$\rho$ with $z$ (see equation \ref{eq:rhoinitial}), must tend to infinity
as $z \rightarrow \infty$ .
The displacements in the plane of the disc
of an infinitely stretched perturbation should effectively vanish, so
$\delta B_r$ and $\delta B_{\phi}$ should be zero at infinity.
This gives us the remaining two boundary conditions required to integrate
the system of equations (\ref{eq:matrix}). Consistently with them, both
$\dE'$ and $\dJ$ vanish as well.

Interestingly, this solution is consistent with $\dE$ and $\dv$ being
non-zero at infinity. The only requirement is that the gradient of the
velocity in the vertical direction $\partial \dv/\partial z$ be zero when $z
\rightarrow \infty$, to prevent any horizontal stretching of the magnetic
field. It may seem puzzling
at first that $\dv$ is non-vanishing at infinity. This can be understood
taking
into account that these perturbations travel to infinity in a finite time
$t_{\infty}$ given by,

\begin{equation}
t_{\infty} = \int_{0}^{\infty} \frac{dz}{v_{Al}} = \frac{H}{v_A}
\int_{0}^{\infty} exp \left( -\frac{z^2}{4} \right) dz = \sqrt{\pi}
\frac{H}{v_A} \,,
	\label{eq:travel_time}
\end{equation}

where $z$ is the vertical coordinate in units of the scaleheight $H$ and
$v_{Al}$ is
the local value of the Alfv\'en speed. Because of this finite travel time to
infinity, the
fluid can retain a finite velocity when $z \rightarrow \infty$. Furthermore,
through equations (\ref{eq:E'phi_v}) and (\ref{eq:E'r_v}) it is clear that
$\dE$, the perturbations in the electric field as seen from the laboratory
frame, are finite at infinity as well.

These boundary conditions are strictly valid at infinity, but will also hold
at a boundary located sufficiently high
above the midplane. We chose to locate the boundary at $z/H = 5$ after
confirming that increasing this height does not significantly
affect either the structure or the growth rate of unstable modes. This can be
appreciated in Table \ref{table:boundary}, which compares the maximum growth
rate $\nu_{max}$ and number of nodes $N$ (proxy for wavenumber) of the
perturbations in all conductivity
regimes for two different locations of the boundary. Summarising,
the boundary conditions adopted at the surface are:

\begin{displaymath}
\dB_r = \dB_{\phi} = 0 \,,
%	\label{eq:bcond}
\end{displaymath}

\noindent
at $z/H = 5$.
\end{description}

This system of equations is solved as a two-point boundary value problem
for coupled ODE by `shooting' from the midplane to the surface of the disc
and simultaneously adjusting the growth rate $\nu$ and $\dE'_{\phi}$ until
the solution converges.

\begin{table}
\caption{Comparison of maximum growth rate $\nu_{max}$ and number of nodes
$N$ of the fastest growing modes for all conductivity regimes and two
different locations of the boundary. In all cases $v_A/c_s=0.1$ and
$\chi_o=10$.}
\begin{tabular}{|l|c|c|c|c|}
\hline
 &$z/H=5$ & & $z/H=7$ & \\
Conductivity Regime & $\nu_{max}$ & $N$ & $\nu_{max}$ & $N$ \\
\hline
Ambipolar Diffusion & $0.7303865$ & $5$ & $0.7303869$ & $5$ \\
Hall limit ($\sigma_1 B_z>0$) & $0.7498761$ & $5$ & $0.7498761$ & $5$ \\
Hall limit ($\sigma_1 B_z<0$) & $0.7461857$ & $5$ & $0.7461854$ & $5$ \\
Comparable conductivities & $0.7345455$ & $5$ & $0.7345459$ & $5$ \\
Opposite conductivities & $0.7340035$ & $5$ & $0.7340039$ & $5$ \\
\hline
\end{tabular}
%\medskip

	\label{table:boundary}
\end{table}

\section{RESULTS}
\label{sec:results}

\subsection{Test Models}
	\label{subsec:test}

We solved the system of equations (\ref{eq:matrix}) for different
conductivity regimes, coupling between fluid components and
initial magnetic field strengths. As discussed in section
\ref{subsec:linearised}, under the linear approximation and disc model
adopted in the present paper, the ambipolar diffusion and resistive
conductivity regimes are identical, so even though throughout
this work we have labelled the case when $\sigma_1=0$ as the
`ambipolar diffusion' limit, it should be kept in mind that this
condition describes the resistive regime as well. Two different
Hall limits exist, as the growth rate of the MRI
depends on the orientation of the initial magnetic field with respect to the
disc angular velocity vector $\bmath{\Omega}$ (W99). The case when $\B_o$ is
parallel (antiparallel) to $\bmath{\Omega}$ is characterised by
$\sigma_1 B_z>0$ ($\sigma_1 B_z<0$).

We calculated the growth rate and vertical structure of all unstable
perturbations for
different conductivity regimes with $v_A/c_s=0.1$. The degree of coupling
between the magnetic field and the neutral component of the fluid was
characterised by either $\chi_o=10$
(good coupling) or $\chi_o=2$ or $1$ (poor coupling). The choice of
$\chi_o$ for
the low coupling analysis is dependent on the conductivity regime of the
fluid. We took $\chi_o=1$ for all regimes, except the Hall ($\sigma_1B_z<0$)
limit, where $\chi_o=2$ was adopted as our code fails to converge
for $\chi<2$. In this regime, the local analysis shows that when $0.5 < \chi <
2$ all wavenumbers grow (W99). %Moreover, in the region
%$1.25 < \chi < 2$ there is still a fastest growing mode, but otherwise the
%growth rate increases monotonically with $k$.
We believe that this complex
structure of the perturbations at ever increasing $k$ prevents our
code from converging when $\chi_o<2$.

We also examined the dependency  of the structure of the fastest growing
modes, their growth rate and the height of maximum amplitude,
with the coupling $\chi_o$  and the strength of the magnetic field for
all conductivity regimes. To study the effect of the magnetic coupling, 
the value of
$v_A/c_s$ was fixed at $0.1$. The impact of the strength of the
field was explored for good and poor coupling conditions.

Finally, we studied the dependency of the growth rate of the most unstable
perturbations with the coupling $\chi_o$ for $v_A/c_s = 0.1$ and $0.01$ and
with the magnetic field strength for $\chi_o = 10$, $2$ and $0.1$.

The relative values of $\sigma_1$ and $\sigma_2$ used to characterise each
conductivity regime, together with the values of $v_A/c_s$ and $\chi_o$
used to
explore the structure of the perturbations are summarised in
Table \ref{table:test_cases}.

\begin{table}
\caption{Relative values of the components of the conductivity tensor
perpendicular to the magnetic field $\sigma_1$ and $\sigma_2$ and fiducial
values of the coupling parameters $\chi_o$ (good coupling, poor coupling
limits) and $v_A/c_s$ adopted to explore the structure of the perturbations
for all conductivity regimes.}
\begin{tabular}{|l|l|l|l|}
\hline
Conductivity Regime & $\sigma$ & $\chi_o$ & $v_A/c_s$  \\
\hline
%\hline
Ambipolar Diffusion & $\sigma_1 = 0$ & $10$, $1$ & $0.1$ \\
%\hline
Hall Limit $\sigma_1 B_z>0$ &$\sigma_2=0$, $\sigma_1>0$ & $10$, $1$ & $0.1$ \\
%\hline
Hall Limit $\sigma_1 B_z<0$ & $\sigma_2=0$, $\sigma_1<0$ & $10$, $2$ &
$0.1$ \\
%\hline
Comparable conductivities & $\sigma_1 = \sigma_2$ & $10$, $1$ & $0.1$ \\
%\hline
Opposite conductivities & $\sigma_1 = -\sigma_2$ & $10$, $1$ & $0.1$  \\
\hline
\end{tabular}
	\label{table:test_cases}
\end{table}

\subsection{Comparison with local analysis}
	\label{subsec:comparison}

\begin{figure} %\subsubsection{fig:unst_modes_amb}
\centerline{\epsfxsize=8cm \epsfbox{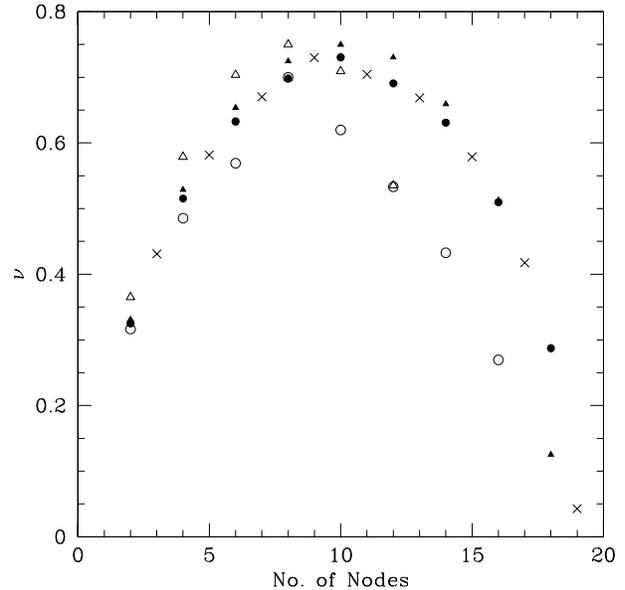}}\vskip 0cm
\caption{Growth rate versus number of nodes (proxy for wavenumber) of
the MRI for different conductivity regimes and coupling at the midplane
$\chi_o$.
Circles show the ambipolar diffusion limit ($\sigma_1 = 0$) and
triangles the Hall limit ($\sigma_2=0$, $\sigma_1 B_z > 0$). Filled symbols
correspond to the good coupling case $\chi_o = 10$ and
open ones to the poor coupling case $\chi_o = 1$. Crosses show the
ambipolar diffusion limit with `even' boundary conditions applied to
$\delta B_r$ and $\delta B_{\phi}$. Note that results in this case are
displaced a quarter of a wavelength (one node) from those obtained with
`odd' boundary conditions. In all cases $v_A/c_s = 0.1$.}
\label{fig:unst_modes_amb}
\end{figure}

The linear growth of the MRI as a function of wavenumber in a local
analysis shows that $\nu$
vs $k$ generally takes the form of inverted quadratics (W99).
The local wavenumber of the perturbations change with $z$,
so we use the number of nodes of $\delta B_r$ over the entire thickness of
the disc, from $z=-5$ to $z=+5$, as a proxy for the wavenumber $k$ to compare
our results with those of W99.
Results are shown in Fig. \ref{fig:unst_modes_amb} for the ambipolar
diffusion and Hall ($\sigma_1 B_z>0$) limits for good and poor coupling. In
both regimes the reduction of the wavenumber of the fastest growing
perturbation with $\chi_o$ is obtained, as expected from the local analysis.
Reducing $\chi_o$ also diminishes the
growth rate of the instability in the ambipolar diffusion limit, but $\nu$
remains unchanged for the Hall regime, as expected from the local results.

These results confirm our expectation that applying boundary conditions
and integrating the fluid equations in the vertical direction would restrict
the unstable frequencies from the continuous curve $\nu$ vs $k$ obtained in
the local analysis to a discrete subset of global unstable modes supported by
the fluid.

Also shown as crosses in Fig.  \ref{fig:unst_modes_amb} are the growth
rates obtained with `even' boundary conditions applied to $\delta B_r$
and $\delta B_{\phi}$ at the midplane (see section \ref{sec:boundary
conditions}), for the ambipolar diffusion limit with $\chi_o=10$.  As
expected, the perturbations are displaced a quarter of a wavelength
(one node) from those obtained with `odd' boundary conditions.

\subsection{Structure of the Perturbations}
	\label{subsec:structure}

\begin{figure} %\subsubsection{fig:B-n plane}
\centerline{\epsfxsize=8cm \epsfbox{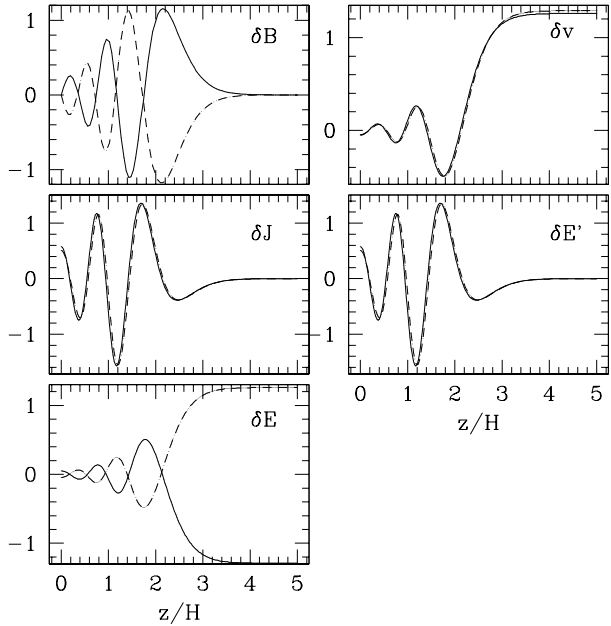}}\vskip 0cm
\caption {Structure of the perturbations in all fluid variables as a function
of height for the
most unstable mode in the ambipolar diffusion limit, for good coupling
($\chi_o=10$) and $v_A/c_s = 0.1$. The growth rate
is $\nu = 0.7304$. Note the non-zero values of $\dv$ and $\dE'$
at the surface, due to the finite travel time to infinity of the
perturbations.}
\label{fig:b_conditions}
\end{figure}

Fig.  \ref{fig:b_conditions} shows the perturbations in all fluid
variables as a function of height, from the midplane ($z/H = 0$) to
the surface of the disc ($z/H = 5$), for the ambipolar diffusion
regime and good coupling ($\chi_o=10$).  They are obtained through
equations (\ref{eq:matrix_JE}) - (\ref{eq:matrix_EJ}) once the ODE
system (\ref{eq:matrix}) has been integrated.  Note the non-zero
values of $\dv$ and $\dE$ at the surface, as discussed in section
\ref{sec:boundary conditions}.  From this point onwards, discussion
will be focussed in the perturbations of the magnetic field only.
Unless otherwise stated, the radial component of the field $\delta
B_r$ is plotted with a solid line and the azimuthal component $\delta
B_{\phi}$ with a dashed line.  As the overall scale of our linear
equations is arbitrary, plots depicting the structure of the
perturbations either do not show the scale of the vertical axis
(corresponding to the amplitude of the perturbations) or show a
conveniently normalised scale, for reference purposes.

\subsubsection{Effect of the Conductivity Regime}
	\label{subsubsec:conductivity}

\begin{figure}
\centerline{\epsfxsize=8cm \epsfbox{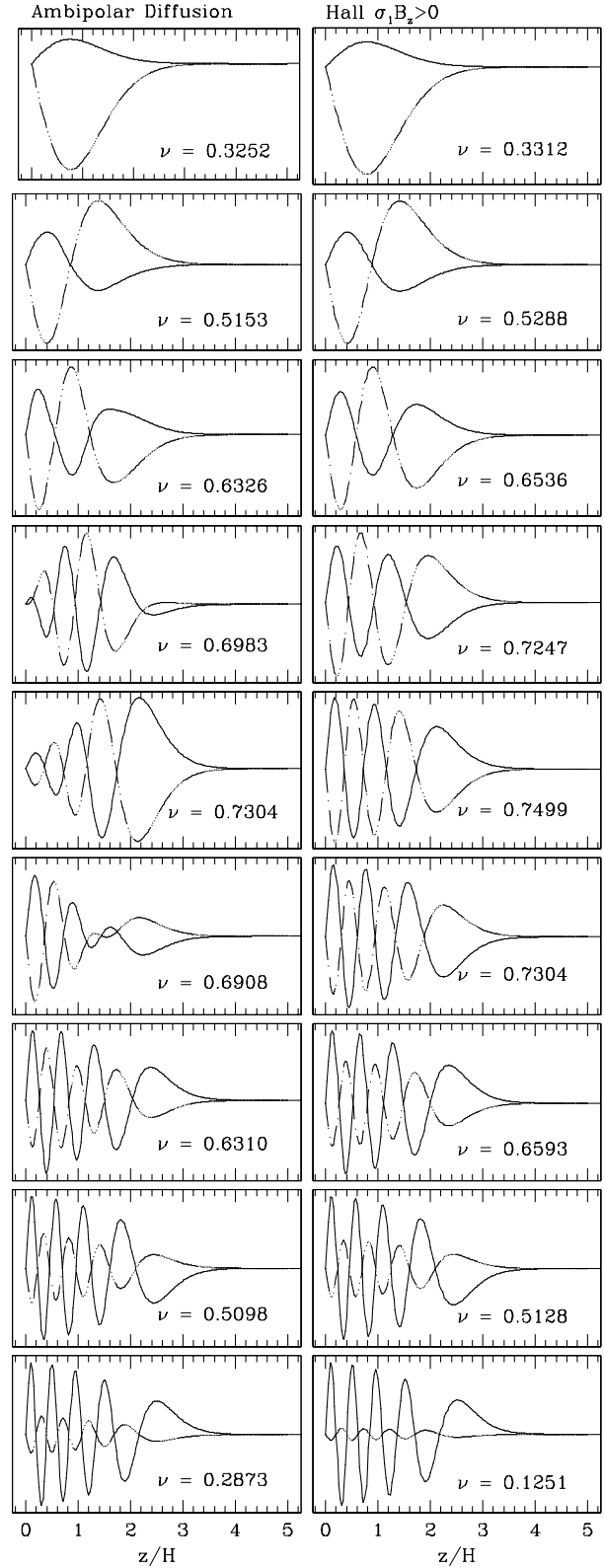}}\vskip 0cm
\caption{Structure and growth rate of all unstable modes of the MRI for the
ambipolar diffusion ($\sigma_1=0$) and Hall ($\sigma_2=0$, $\sigma_1 B_z >0$)
cases. In all plots $\chi_o=10$ and $v_A/c_s = 0.1$. For this good coupling,
there are no differences between both Hall limits. When both conductivity
components are present ($\sigma_1 = \pm \sigma_2$ cases), results resemble the
ambipolar diffusion limit shown.}
\label{fig:struct_comp}  %\subsubsection{fig:MRN conductivity}
\end{figure}

Fig. \ref{fig:struct_comp} compares the structure of all unstable
perturbations for the ambipolar diffusion and Hall ($\sigma_1 B_z>0$) regimes
under good coupling ($\chi_o = 10$). The modes are ordered by the number of
nodes. We find that at small $\nu$, the structure of the perturbations in
both cases is very similar, but significant differences arise when the growth
rate is close to maximum. Then, ambipolar diffusion perturbations peak at the
node closest to the surface, while Hall ($\sigma_1 B_z > 0$) ones peak
closest to the midplane. This behaviour is linked to the change in the local
coupling $\chi$ with $z$ and its effect in the structure of the
perturbations for different conductivity regimes. It will be discussed
further in the next section. At this good coupling level,
there are no appreciable differences in the structure or growth rate of
perturbations between both Hall limits results, as expected from the local
analysis (W99). When both conductivity components are present
($\sigma_1 = \pm \sigma_2$ cases), the structure of the perturbations is
similar to
the ambipolar diffusion limit. This property is also dependent on the value of
the local coupling $\chi$ and will be analysed in the next section.
We found $9$ to $10$ unstable perturbations in all cases.

Under low coupling conditions, $\chi_o = 2$ or $1$ depending on the
conductivity regime (see Table \ref{table:test_cases}), fewer unstable modes
grow for both the ambipolar diffusion, Hall ($\sigma_1 B_z>0$) and the
comparable conductivities ($\sigma_1 = \sigma_2$) regimes.
Six to eight unstable perturbations are found in these cases. As expected from
the local analysis (W99), the range of wavenumbers for which unstable modes
exist is reduced as compared with the good coupling cases.

\begin{figure}
\centerline{\epsfxsize=7.5cm \epsfbox{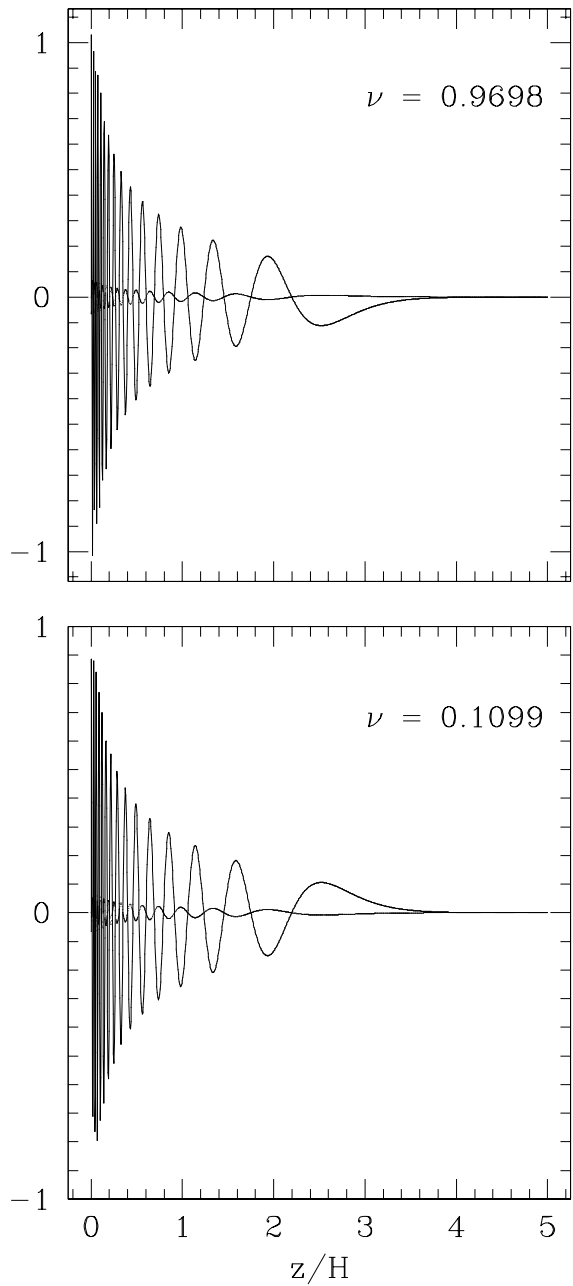}}\vskip 0cm
\caption{Structure of two unstable modes in the Hall ($\sigma_1 B_z< 0$) limit
for poor coupling ($\chi_o = 2$) and $v_A/c_s=0.1$. Note the
complex structure of the perturbations (high wavenumber). Unstable modes are 
so closely spaced that increasing the number of
nodes only marginally changes their wavenumber and growth rate.}
\label{fig:struct_hall_p1}  %\subsubsection{fig:MRN conductivity}
\end{figure}

Results are
quite different for the two remaining conductivity regimes (Hall
$\sigma_1B_z<0$ and $\sigma_1=-\sigma_2$). More unstable modes are
found in these
cases; $12$ in the opposite conductivities case ($\sigma_1 = -\sigma_2$) and a
total of $27$ for the Hall ($\sigma_1 B_z <0$) limit.
In this last case in particular, despite the low coupling of ionised and
neutral components of the fluid, unstable modes have a very complex structure
(high wavenumber). Fig.\ \ref{fig:struct_hall_p1} shows the
structure of two such modes at the low growth rate, high wavenumber
region of the $\nu$ - $k$ space. Note that
unstable modes are so closely spaced that increasing the number of nodes by
a few only maginally changes their wavenumber and has little effect on their
growth rate. This complexity is expected from
the form of the dispersion relation at low coupling from the local analysis
(W99). Non-linear simulations (Sano \& Stone 2002a, b) confirm that the many
growing modes in this regime strongly interact with each other and the
instability develops into MHD turbulence. This turbulence is a transient
phase that eventually dies away in two-dimensional simulations (Sano \& Stone
2002a) but it is sustained in full three-dimensional models (Sano \& Stone
2002b). In both cases the non-emergence of the typical two-channel flow
obtained in other regimes is noted by the authors.

\begin{figure}
\centerline{\epsfxsize=8cm \epsfbox{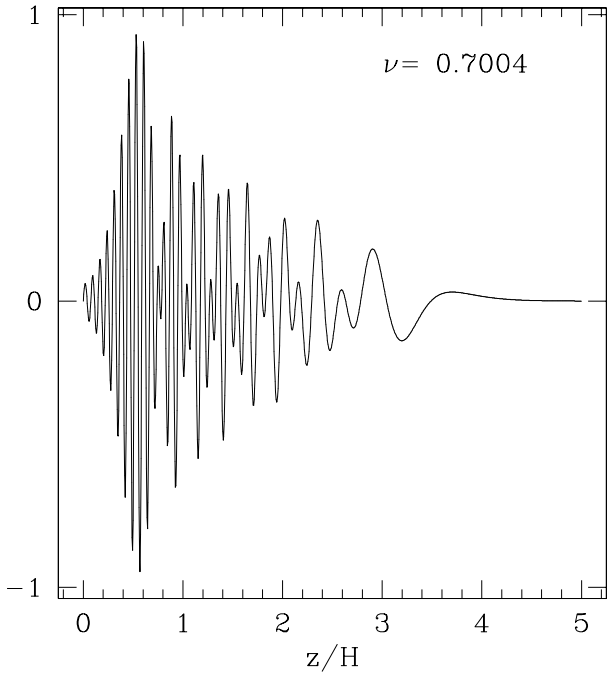}}\vskip 0cm
\caption{Structure of an unstable mode in the ambipolar diffusion limit
($\sigma_1=0$) for good coupling ($\chi_o = 10$) and $v_A/c_s=0.01$.
Note the interference pattern of the perturbation, which
suggests that this global mode is a superposition of two WKB modes of similar
growth rates.}
\label{fig:interference}  %\subsubsection{fig:MRN conductivity}
\end{figure}

Finally, some of the perturbations show a structure resembling an
interference pattern (see Fig.\ \ref{fig:interference}).  They were
obtained specifically in regimes where ambipolar diffusion is present,
for good and poor coupling, but not in any of the Hall limits.
This pattern can be explained recalling that local results show that
two unstable modes exist with the same growth rate and different
wavenumber.  Despite this, just one global mode is found for each
$\nu$ in this analysis.  Again, the application of boundary conditions
and integration along the vertical direction restricts global unstable
modes from those possible under a local analysis.  The interference
pattern suggests that global modes are a superposition of two WKB
modes with (nearly) the same growth rate and which are not global
solutions themselves.  The interference of Fig.
\ref{fig:interference} was successfully replicated through the
superposition of two local modes with $\nu = 0.7004$ using the
analytical expresions in W99 for the ambipolar diffusion limit.

\subsubsection{Effect of the coupling parameter $\chi_o$}
	\label{subsubsec:coupling}

\begin{figure*}
\centerline{\epsfxsize=17cm \epsfbox{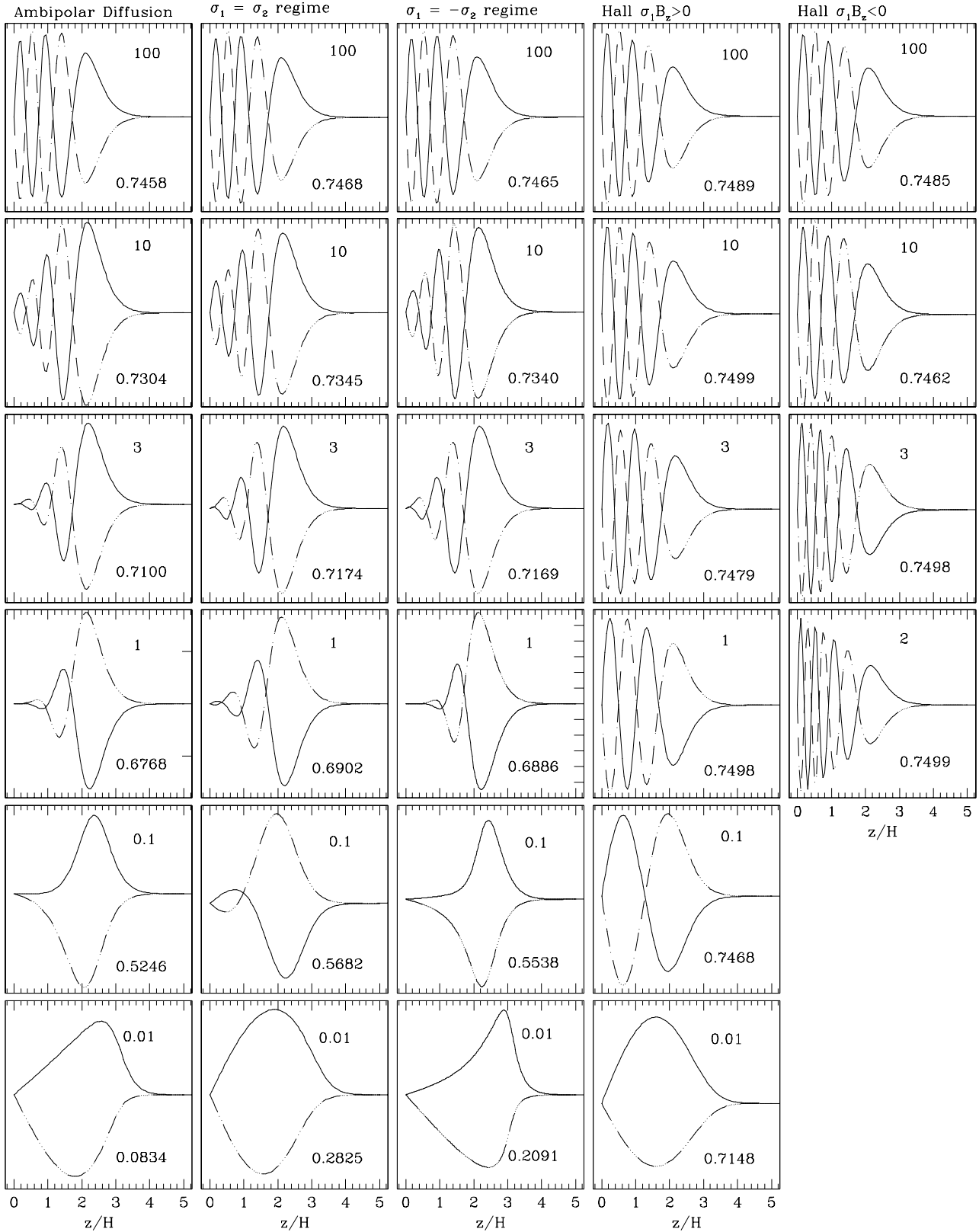}}\vskip 0cm
\caption{Structure and growth rate of the most unstable modes of the MRI for
all conductivity regimes and different values of $\chi_o$. In all cases
$v_A/c_s = 0.1$. The value of the coupling parameter $\chi_o$ is
indicated at the top right corner of each panel. The growth rate ($\nu$) of
the
perturbations is shown in the lower right corner. Note that the Hall $\sigma_1
B_z<0$ regime is explored for $\chi_o \ge 2$. In this conductivity
regime our code fails to converge for $\chi<2$ as in this region of parameter
space the range of wavenumbers for which local unstable modes exist
becomes infinite (W99).}
\label{fig:amb_hall_modes}  %\subsubsection{fig:MRN conductivity}
\end{figure*}

Fig.\ \ref{fig:amb_hall_modes} compares the structure and growth rate
of the most unstable modes of the MRI for all conductivity regimes as
a function of the coupling parameter $\chi_o$.  In all cases
$v_A/c_s=0.1$.  We notice that reducing the coupling $\chi_o$ causes
the wavenumber (i.e.  the number of nodes) of unstable modes to
diminish in all conductivity regimes except the Hall $\sigma_1 B_z<0$
limit (rightmost column of Fig.  \ref{fig:amb_hall_modes}), for which
this dependency is inverted.  The growth rate is also reduced at a
rate that depends on the conductivity regime of the fluid.  These
results are expected from the findings of the local analysis (W99).

\begin{figure}
\centerline{\epsfxsize=8cm \epsfbox{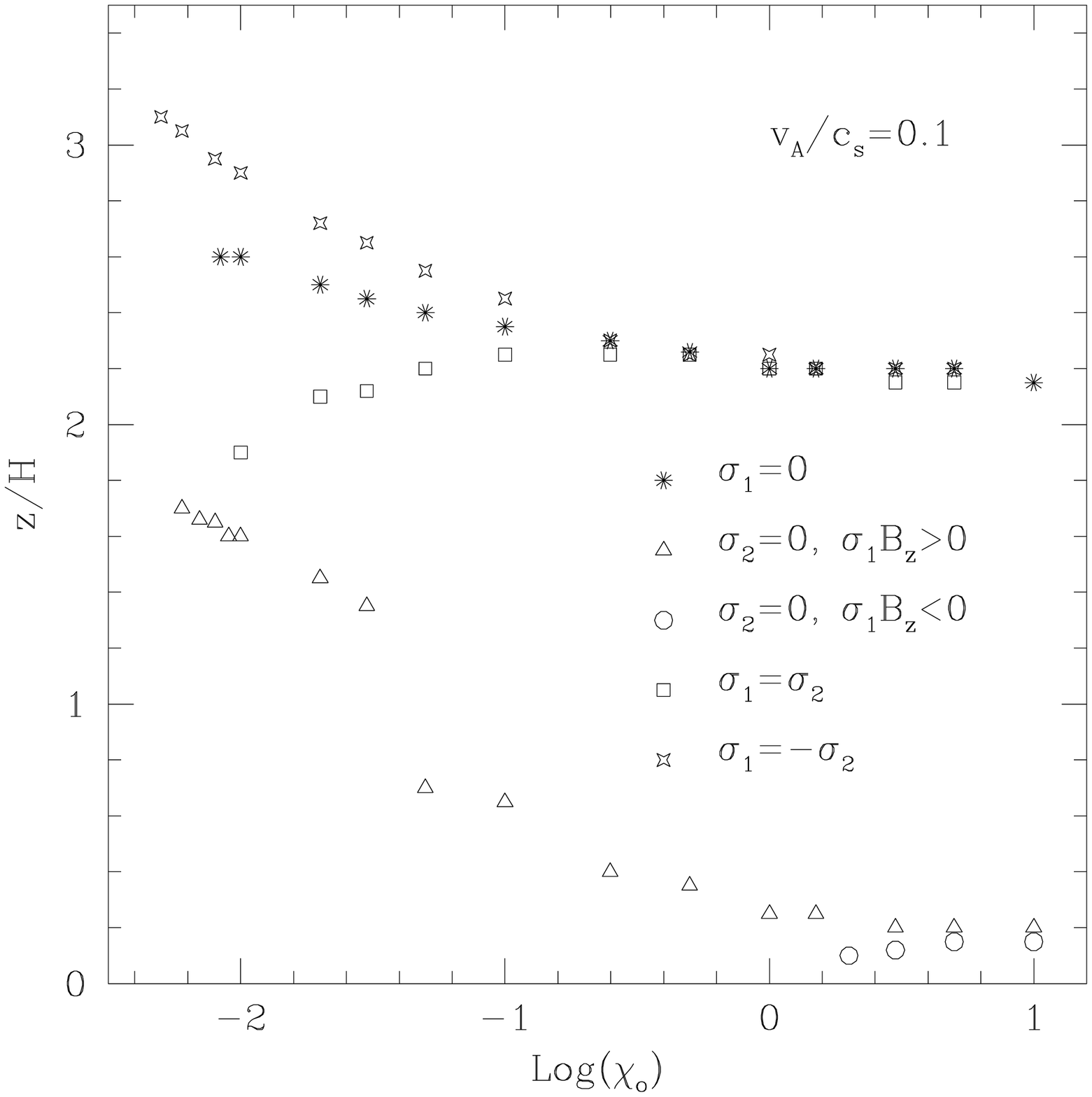}}\vskip 0cm
\caption{Height of maximum amplitude of the most unstable perturbations as a
function of $\chi_o$ for all conductivity regimes. In all cases
$v_A/c_s = 0.1$}
\label{fig:zeta}  %\subsubsection{fig:MRN conductivity}
\end{figure}

It is evident from Fig.\  \ref{fig:amb_hall_modes} that at very high
magnetic coupling ($\chi_o \approx 100$), the fluid is close to ideal
MHD conditions and results obtained in all conductivity regimes are
alike.  When the coupling is reduced to $\chi_o \approx 10$, we begin to
appreciate differences between them.  In particular, the amplitude of
the perturbations when ambipolar diffusion is present peaks close to
the surface while in both Hall limits the maximum amplitude is closest
to the midplane.  This is more clearly appreciated in Fig.
\ref{fig:zeta}, which plots the height of maximum amplitude of the
fastest growing modes as a function of the coupling parameter $\chi_o$
and the conductivity regime of the fluid for $v_A/c_s = 0.1$.  This
figure shows that pure Hall regimes ($\sigma_1 B_z >0$ and $\sigma_1
B_z<0$) peak closer to the midplane, for all $\chi_o$ studied, than
the cases when ambipolar diffusion is present.  This behaviour can be
explained by the dependency of the local growth of the instability
with $\chi$ for different conductivity regimes.  The maximum growth
rate of ambipolar diffusion perturbations increases with the local
$\chi$ (W99), which in turn is a function of height.  As a result, at
higher $z$, the local growth of the instability increases, driving the
amplitude of global perturbations to increase.  Hall ($\sigma_1
B_z>0$) perturbations, on the contrary, have the same $\nu_{max}$ for
all $\chi$, so the instability is not driven from any particular
vertical location, which explains the flatness of their envelope.
This also explains why in Fig.  \ref{fig:struct_comp} the differences
between these regimes are apparent at close to maximum growth: The
increment in the local growth rate with the coupling $\chi$ is less
marked for slow growing perturbations.  This is also appreciated in
the local analysis by the form of the dispersion relation for
different $\chi$ in the ambipolar diffusion regime (W99).

It is also clear from Fig. \ref{fig:zeta} that perturbations in the
ambipolar
diffusion and Hall ($\sigma_1B_z>0$) limits peak at higher $z$ when $\chi_o$
is reduced. This dependency is driven by the reduction of the
wavenumber of most unstable perturbations with the coupling in these regimes
(W99, see also Fig. \ref{fig:amb_hall_modes}).
On the contrary, in the Hall ($\sigma_1B_z<0$) limit, the wavenumber of the
fastest growing mode increases as
$\chi_o$ is reduced and the perturbations peak closer to the midplane with
weaker $\chi_o$.

Finally, looking at the first three columns of Fig.
\ref{fig:amb_hall_modes} it is evident that the structure of unstable modes
in the
$\sigma_1 = \pm \sigma_2$ conductivity regimes are remarkably similar to the
ambipolar diffusion limit shown in the leftmost column of the figure
for $\chi_o \gtrsim v_A/c_s = 0.1$. When the coupling is weaker than
this value, the structure of unstable modes in these regimes is no longer
alike. To explain this we recall that
the minimum degree of coupling for unstable modes to grow, determined by
the requirement that a wavelength fit in the disc scaleheight, is given by
$\chi \gtrsim v_A/c_s$ in the ambipolar
diffusion limit, and $\chi \gtrsim {v_A}^2/{c_s}^2$ in the Hall case (W99).
For $\chi_o \lesssim 0.1$ then, the growth rate of ambipolar diffusion
($\sigma_1= 0$)
perturbations is expected to drop markedly and the envelope of the
perturbations will be mainly determined by the Hall effect. %(see section
%\ref{subsubsec:parameter_coupling} for analysis of the growth rate of
%perturbations as a function of $\chi_o$).
This transition in both
$\sigma_1 = \pm \sigma_2$ cases, is clearly seen in Fig. \ref{fig:zeta}.
In both cases, for $\chi_o \gtrsim 0.1$ the perturbations
resemble the ambipolar diffusion limit ($\sigma_1=0$).
In the $\sigma_1 = \sigma_2$ regime, when
$\chi_o < v_A/c_s$ the perturbations resemble those in the
Hall ($\sigma_1 B_z>0$) limit (compare lowest panels in the second and fourth
columns of Fig. \ref{fig:amb_hall_modes}), consistent with the notion that
ambipolar diffusion effects are no longer important in this region of
parameter
space. This
implies a constant $\nu_{max}$ for weaker $\chi_o$, so the perturbations tend
to peak closer to the midplane. On the contrary, in the $\sigma_1=-\sigma_2$
case, $\nu_{max}$ continues to diminish with $\chi$ once $\chi<v_A/c_s$ (W99).
In this case, the range of growing modes is always finite (as opposed to the
Hall $\sigma_1 B_z<0$ limit) and there is a fastest growing mode for every
$\chi$. As a result, the instability is driven at intermediate $z$ (as
in the ambipolar diffusion limit) and accordingly, the perturbations peak at
higher $z$
with decreasing $\chi_o$. The height of maximum amplitude increases
faster as $\chi_o$ is reduced in this regime than in the ambipolar diffusion
limit. This occurs because $\nu_{max}$ for a given local $\chi$ is greater in
this case (W99, see also section \ref{subsubsec:parameter_coupling}) so
global perturbations are amplified even further.

\subsubsection{Effect of the Magnetic Field Strength}
	\label{subsubsec:magnetic_field}

\begin{figure}
\centerline{\epsfxsize=8cm \epsfbox{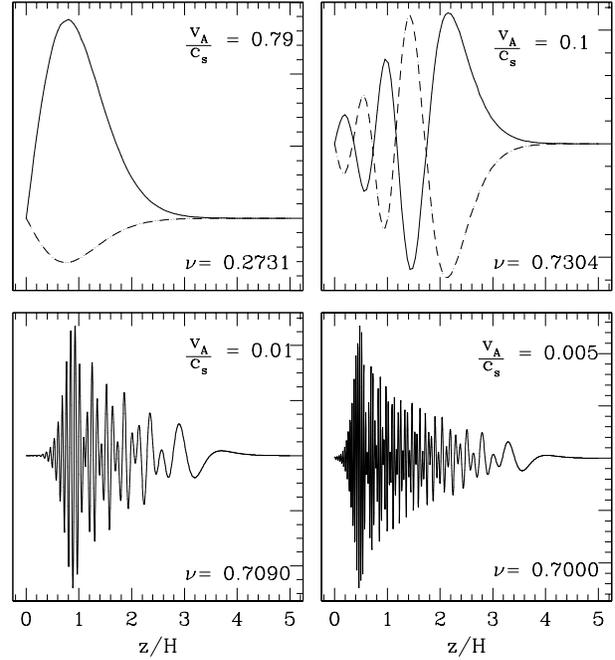}}\vskip 0cm
\caption{Structure as a function of height and growth rate of the MRI
for different choices of $v_A/c_s$ under the ambipolar diffusion limit
($\sigma_1 = 0$) and good coupling $\chi_o=10$. The value of $v_A/c_s$ is
indicated at the top right corner of each panel. The growth rate is shown in
the lower right corner.}
%Clockwise from top left panel,
%$v_A/c_s=0.79$, $0.1$, $0.005$ and $0.01$.
\label{fig:field_amb} %\subsubsection{fig:MRN contributions}
\end{figure}

\begin{figure}
\centerline{\epsfxsize=8cm \epsfbox{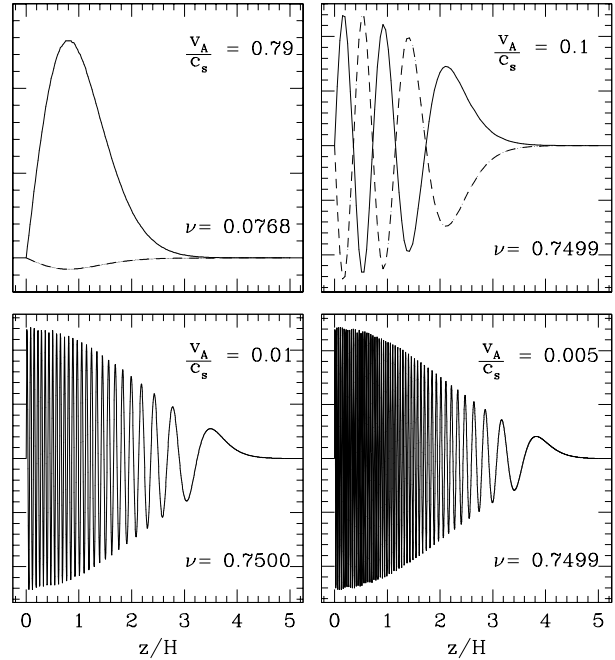}}\vskip 0cm
\caption{As for Fig. \ref{fig:field_amb}, but for the Hall limit $\sigma_1
B_z>0$. The Hall limit when $\sigma_1 B_z<0$ (not shown) exhibits the same
dependency with $v_A/c_s$ for this $\chi_o$.}
\label{fig:field_hall} %\subsubsection{fig:MRN contributions}
\end{figure}

In ideal MHD, the weaker the magnetic field the higher the
minimum wavenumber  of the perturbations (Balbus \& Hawley 1991). The
results of this study are consistent with this finding. With $v_A/c_s
\sim 0.005$ the perturbations grow with very high wavenumbers in all
conductivity regimes. This can
be appreciated in Fig. \ref{fig:field_amb} and \ref{fig:field_hall} for the
ambipolar diffusion and Hall ($\sigma_1 B_z>0$) limits, respectively, for
good coupling ($\chi_o=10$). At this $\chi_o$, solutions for both Hall
limits are
similar. Also, $\sigma_1 = \pm \sigma_2$ regimes are similar to the ambipolar
diffusion limit, as expected. Note the interference pattern of perturbations
in lower panels of Fig. \ref{fig:field_amb}.

\begin{figure}
\centerline{\epsfxsize=8cm \epsfbox{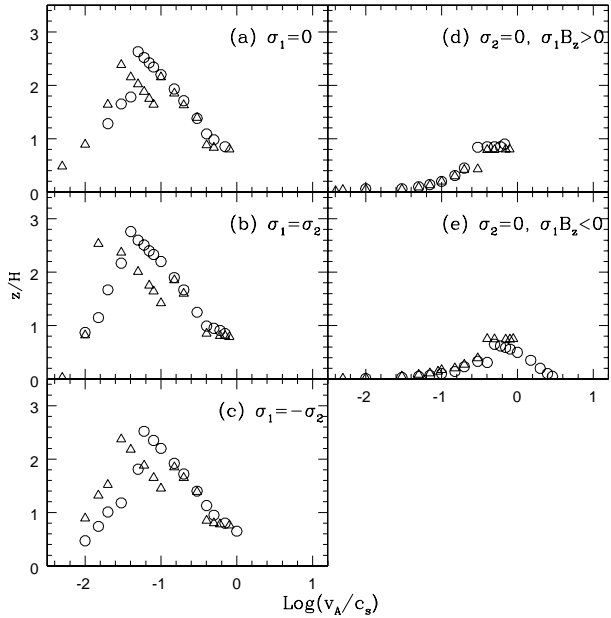}}\vskip 0cm
\caption{Height of maximum amplitude of the perturbations as a function of
$v_A/c_s$, for all conductivity regimes. Triangles correspond to good
coupling ($\chi_o=10$) and circles to poor coupling ($\chi_o=2$).}
\label{fig:zeta_field}  %\subsubsection{fig:MRN conductivity}
\end{figure}

We also studied the dependency of the height of maximum amplitude with the
strength of the magnetic field (Fig. \ref{fig:zeta_field}) for good
($\chi_o=10$) and poor coupling ($\chi_o=2$).
In the former case, when ambipolar diffusion is present, with and
without the Hall effect and regardless of the sign of $\sigma_1 B_z$, the
location of maximum amplitude of the fastest growing
perturbations as a function of $v_A/c_s$ is similar, which is expected as
$\chi_o > v_A/c_s$ (see Fig. \ref{fig:zeta_field}, panels
$a$, $b$ and $c$). We obtained unstable modes for $v_A/c_s$ up to $1$. As
$v_A/c_s$ is reduced from this value, the perturbations peak at higher $z$
until $v_A/c_s
\sim 0.04$. %This is due to the combined effect of the increase of wavenumber
However, the location of maximum amplitude begins
to diminish as the field is further reduced, which could be caused by the
interference pattern, and very high wavenumber, of the perturbations (see
also Fig.
\ref{fig:field_amb}). The height of maximum amplitude peaks at $z \sim 2.8$.

In both Hall limits the height of
maximum amplitude increases with the strength of the magnetic field until
$v_A/c_s \sim 0.3$ and then remains unaffected with further increments of
$v_A/c_s$ (panels $d$ and $e$ of Fig. \ref{fig:zeta_field}). This occurs
because as the strength of the
magnetic field increases, the wavenumber of the perturbations diminish, which
pushes the
maximum amplitude to higher $z$. For $v_A/c_s \sim 0.3$ the
perturbations have only one node, so any further increase in the
magnetic field strength have little effect in the location of the maximum
amplitude.

In the low coupling case results are very similar to the $\chi_o=10$ cases. We
note that in regimes where ambipolar diffusion is present (left hand side
panels
of Fig. \ref{fig:zeta_field}), perturbations tend to peak at a lower $z/H$
than in
the good coupling cases when $v_A/c_s \lesssim 0.1$. In the
Hall ($\sigma_1 B_z<0$) regime, there are unstable modes for `suprathermal'
field
strengths ($v_A/c_s$ up to $2.9$). This will be further analysed in section
\ref{subsubsec:parameter_field}, dealing with the dependency of the growth
rate of this instability with the strength of the magnetic field. In this
case, the height of maximum amplitude peaks at $v_A/c_s \approx 0.5$ and then
gradually diminishes as the field is incremented beyond this value.

\subsection{The perturbations in parameter space}
	\label{subsec:parameter_space}

\subsubsection{Effect of the coupling}
	\label{subsubsec:parameter_coupling}

\begin{figure} %\subsubsection{fig:B-n plane}
\centerline{\epsfxsize=8cm \epsfbox{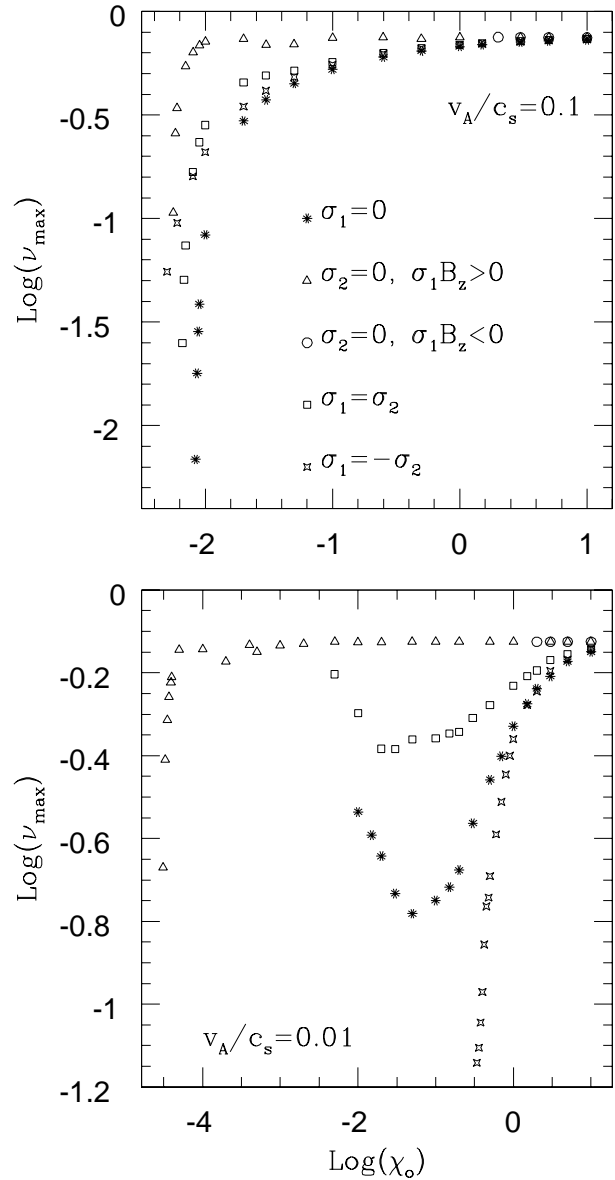}}\vskip 0cm
\caption {Growth rate of the fastest growing modes of the MRI as a function
of $\chi_o$ for different conductivity regimes. $v_A/c_s = 0.1$ (top panel)
and $0.01$ (bottom panel).}
\label{fig:fast_modes_coupling}
\end{figure}

Fig. \ref{fig:fast_modes_coupling} shows the growth rate of the most unstable
modes as a function of the coupling $\chi_o$ for all
conductivity regimes with $v_A/c_s=0.1$ (top panel) and $0.01$ (bottom panel).
The Hall ($\sigma_1 B_z <0$) limit  could not be modelled for
$\chi_o<2$ because in this region of parameter space the range of wavenumbers
for which unstable modes grow becomes infinite (see section
\ref{subsec:test}). We find that at
the good coupling limit
the instability grows at a rate similar to its ideal value of $0.75 \Omega$
for all conductivity regimes. As the coupling diminishes,
the growth rate is reduced at a rate that depends on the conductivity regime
of the fluid.  In the Hall ($\sigma_1 B_z >0$) case the growth rate remains
unaffected until $\chi_o \sim 0.01$, and then
diminishes drastically to $0.1 \Omega$ when $\chi_o \sim 0.005$. The
ambipolar diffusion limit has a much more gradual reduction of $\nu_{max}$
with $\chi_o$. In this case, the growth
rate departs significantly from the ideal value for $\chi_o \sim 0.1$ and then
drops rapidly, reaching $\sim 0.007\Omega$ for $\chi_o \sim 0.008$.
This is in agreement with findings by W99 that unstable modes grow when
$\chi \gtrsim v_A/c_s$ in the ambipolar diffusion limit, and
$\chi \gtrsim {v_A}^2/{c_s}^2$ in the Hall case (see also section
\ref{subsubsec:coupling}). When $\chi$ is less than these values,
perturbations are strongly damped.

When $v_A/c_s \sim 0.01$ (bottom panel of Fig. \ref{fig:fast_modes_coupling}),
the growth of Hall ($\sigma_1B_z>0$) perturbations is constant at about the
ideal rate $0.75\Omega$ until $\chi_o \sim 10^{-4}$. Below this, it
plummets to
zero as expected. On the contrary, in the ambipolar diffusion and
$\sigma_1 = \pm \sigma_2$ regimes, $\nu_{max}$ begins to diminish much sooner.
It is also noticed that, in the ambipolar diffusion and $\sigma_1 =
\sigma_2$ cases, the growth rate increases again after reaching a minimum
for $\chi_o \sim 0.05$. This is caused by the high wavenumber of the
perturbations
due to the weakness of the magnetic field. In these
conditions global effects (stratification) are less important and the maximum
growth rate diminishes with the coupling as per local results (W99). As the
wavenumbers of the perturbations decrease when $\chi_o$ is reduced, for
sufficiently low $\chi_o$ ( $\sim 0.05$), $k$ is low enough for global
effects  to be important again and modify the growth rate of unstable
perturbations. With the local $\chi$ increasing with height
above the midplane, stratification will tend to increase the growth of global
modes at low $\chi_o$. 

\subsubsection{Effect of the magnetic field strength}
	\label{subsubsec:parameter_field}

\begin{figure} %\subsubsection{fig:B-n plane}
\centerline{\epsfxsize=8cm \epsfbox{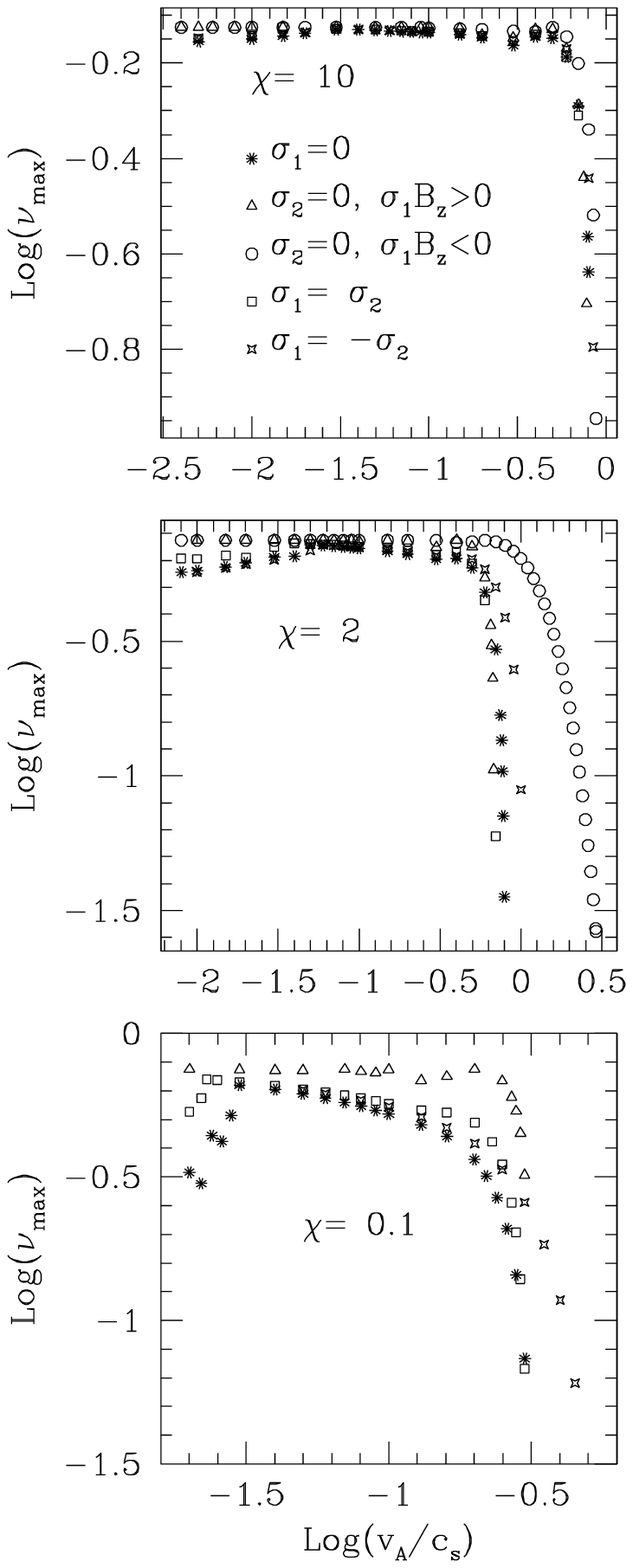}}\vskip 0cm
\caption {Growth rate of the fastest growing modes of the MRI as a function
of $v_A/c_s$ for different conductivity regimes. From top to bottom the
coupling $\chi_o$ is $10$, $2$ and $0.1$.}
\label{fig:fast_modes_field}
\end{figure}

The dependency of the maximum growth rate with the
strength of the magnetic field for all conductivity regimes is shown in Fig.
\ref{fig:fast_modes_field} for $\chi_o=10$, $2$ and $0.1$. In the good
coupling
case (top panel), increasing the strength of the magnetic field has little
effect in the growth rate of the most unstable modes of all conductivity
regimes until $v_A/c_s \sim 1$, where it drastically drops to zero. These
results are similar to the ideal MHD case, which predicts that at this
strength of the magnetic field the wavelength of most unstable modes become
$\sim$ H, the scaleheight of the disc, and the perturbations are strongly
damped.

In the $\chi_o=2$ case shown in the middle panel, we found unstable
modes in the Hall limit ($\sigma_1 B_z <0$) for $v_A/c_s$ up to $2.9$.
We know from the local analysis (W99) that once the local $\chi \le 2$, 
unstable modes exist for every $kv_A/\Omega$ in this conductivity regime. As a
result, even for suprathermal fields ($v_A/c_s>1$), there are still
unstable modes with $kH \lesssim 1$ growing within the disc.

Results for $\chi_o=0.1$ (bottom panel)
show clearly how $\nu_{max}$ plummets when $v_A/c_s \gtrsim \chi$ (ambipolar
diffusion limit) or ${{v_A}^2}/{{c_s}^2} \gtrsim \chi$ (Hall limit), as
expected.

Finally, turning our attention to the dependency of the growth rate of
the instability with the field strength at low $v_A/c_s$, we
appreciate in all panels of Fig.  \ref{fig:fast_modes_field} that
$\nu_{max}$ initially increases as $v_A/c_s$ is reduced, until it
reaches a maximum.  Further reductions in $v_A/c_s$ cause the growth
rate to diminish monotonically.  Comparing these results with the
findings of the local analysis (W99), we can show that the growth
rates of global unstable modes at weak magnetic field strengths tend
to the local values for the relevant coupling $\chi_o$.  This can be
explained simply by the increase in wavenumber of unstable modes as
$v_A/c_s$ is reduced, which causes global effects (driven by
stratification) to be less important.  As a result, the growth of
global modes does not differ significantly from the local values at
the same coupling.

\section{Discussion}
\label{sec:discussion}

The solutions presented in the previous sections illustrate the growth
and structure of the MRI when different components of the conductivity
tensor are dominant throughout the entire cross-section of the disc. Density 
stratification causes the local growth of unstable modes, and the amplitude 
of global perturbations, to be a function of height. The envelopes of 
short-wavelength solutions are shaped by this competition between different 
growth rates acting at different vetical locations. Moreover, when $\chi$ is 
weak at the midplane, long wavelength perturbations are important and 
vertical stratification is crucial in determining the growth of global MRI 
solutions. The results presented in this contribution confirm these 
expectations. 

When the Hall and Pedersen components are comparable, the Hall effect
alters the structure and growth of unstable perturbations for
$\chi_o<v_A/c_s$.  In this region of parameter space, ambipolar
diffusion perturbations have negligible growth, but unstable modes
that include Hall conductivity still grow at $\nu = 0.2 - 0.3$.  Hall
$\sigma_1 B_z>0$ perturbations grow faster than $\sigma_1
B_z<0$ ones.  Furthermore, under a weak magnetic field ($v_A/c_s
\lesssim 0.01$), the Hall effect significantly increases the growth
rate of unstable modes at low coupling.  When it dominates, unstable
modes grow at close to the ideal rate for $\chi_o \sim 10^{-4}$.

The height above the midplane where the most
unstable perturbations peak is dependent on the conductivity regime of
the fluid.  Consequently, the vertical location of the active zones
within the disc, in which the MRI produces angular momentum transport
and disc material is being accreted, is dependent on the configuration
of the conductivity tensor.  Perturbations including ambipolar
diffusion peak consistently higher than those in both Hall limits.
Also, for $\chi_o<v_A/c_s$, $\sigma_1= \pm \sigma_2$ modes peak at
different heights, signalling that the Hall effect is dependent on the
orientation of the magnetic field with respect to the disc angular
velocity vector $\bmath{\Omega}$.  In this region of parameter space,
when $\sigma_1 B_z<0$, modes peak at a higher $z$ than when $\sigma_1
B_z>0$.  The Hall effect does not significantly modify the dependency
of the height of maximum amplitude of unstable modes with the strength
of the magnetic field for good coupling.  In this region of parameter
space ambipolar diffusion dominates and causes the perturbations to
peak at higher $z$ when $v_A/c_s$ is reduced.

When $\sigma_1 B_z<0$, Hall perturbations can have a very complex
structure (high wavenumber), even at low coupling, and many modes are
found to grow.  In the non-linear stage, the interaction between these
modes causes the MRI to develop into MHD turbulence with non-emergence
of the typical two-channel flow obtained in other conductivity regimes
(Sano \& Stone 2002a, 2002b).

When does Hall diffusion determine the behaviour of the instability?
Naively one might propose $|\sigma_1| \ga \sigma_2$ as a criterion,
but this ignores the level of coupling between the magnetic field and
the gas.  For example if $\chi\ga 10$, ideal MHD almost holds and
there is little dependence of growth rate and structure on the
diffusion regime (see Fig.\ \ref{fig:struct_comp}).  A useful
criterion can be derived using the results of the local analysis in
W99 and comparing the maximum growth rates with and without Hall
diffusion in the weak-coupling limit.  In the absence of Hall
diffusion (i.e.\ $\sigma_1 = 0$), the maximum growth rate for $\chi\la
1$ is $3\chi/4$.  When Hall diffusion is present and $\chi \la
|\sigma_1|/ \sigma_\perp $, the maximum growth rate is either
$\frac{3}{4}|\sigma_1|/(\sigma_\perp + \sigma_2)$ if $\sigma_1 B_z >
0$ or the instability is suppressed if $\sigma_1 B_z < 0$.  In either
case, Hall diffusion dominates the behaviour of the instability when

\begin{equation}
%\chi \, \la \, \chi_\mathrm{crit} = \frac{|\sigma_1|}{(\sigma_\perp
%+ \sigma_2)} \,.
\chi \, \la \, \chi_\mathrm{crit} = \frac{|\sigma_1|}{\sigma_\perp} \,.
	\label{eq:chi_crit}
\end{equation}

\noindent
Thus even when $|\sigma_1| \ll \sigma_2$, the structure and growth
rate of the magnetorotational instability are dominated by Hall
diffusion if $\chi < |\sigma_1|/\sigma_{\perp}$.  This is easily
satisfied, for example, for the nominal conditions at the disc
midplane 1 AU from the central protostar (see Table
\ref{table:ionisation}), where $|\sigma_1|/\sigma_{\perp} \approx
|\beta_e| = 0.035$ (using eqs. \ref{eq:sig1}  and \ref{eq:sigperp} with
$|\beta_e|\gg 1$) and $\chi = 9\ee -6 $.  At 5 AU, $\chi\approx 2$ and
$\sigma_1 \gg \sigma_2 $, so Hall diffusion dominates here also.
Although these results depend on the assumed magnetic field strength,
the conditions under which $\chi < \chi_\mathrm{crit}$ are so broad
that we can conclude that Hall diffusion determines the growth rate and
structure of the instability over a large range of radii.

Despite this, Hall diffusion has generally been neglected in studies
of accretion discs in favour of the ambipolar diffusion or resistive
limits.  Here we illustrate the severity of this approximation by
comparing the structure and growth rate of the unstable perturbations
for a model with $\sigma_1 = \sigma_2$ to those for `simplified' pure
ambipolar/resistive diffusion and Hall ($\sigma_1B_z>0$) models
obtained by setting $\sigma_2$ or $\sigma_1$ to zero respectively and
reducing the coupling parameter $\chi_o$ by a factor of $\sqrt{2}$ to
reflect its dependence on $({\sigma_1}^2 + {\sigma_2}^2)^{1/2}$ (see
eq.  \ref{eq:chi}).  The full model has $v_A/c_s = 0.01$ and
$\chi_o=0.01414$, whereas the the corresponding ambipolar diffusion
and Hall limits have $\chi_o = v_A/c_s = 0.01$ and the appropriate
values of $\sigma_1$ and $\sigma_2$ as per Table
\ref{table:test_cases}.

\begin{figure} %\subsubsection{fig:B-n plane}
\centerline{\epsfxsize=7.5cm \epsfbox{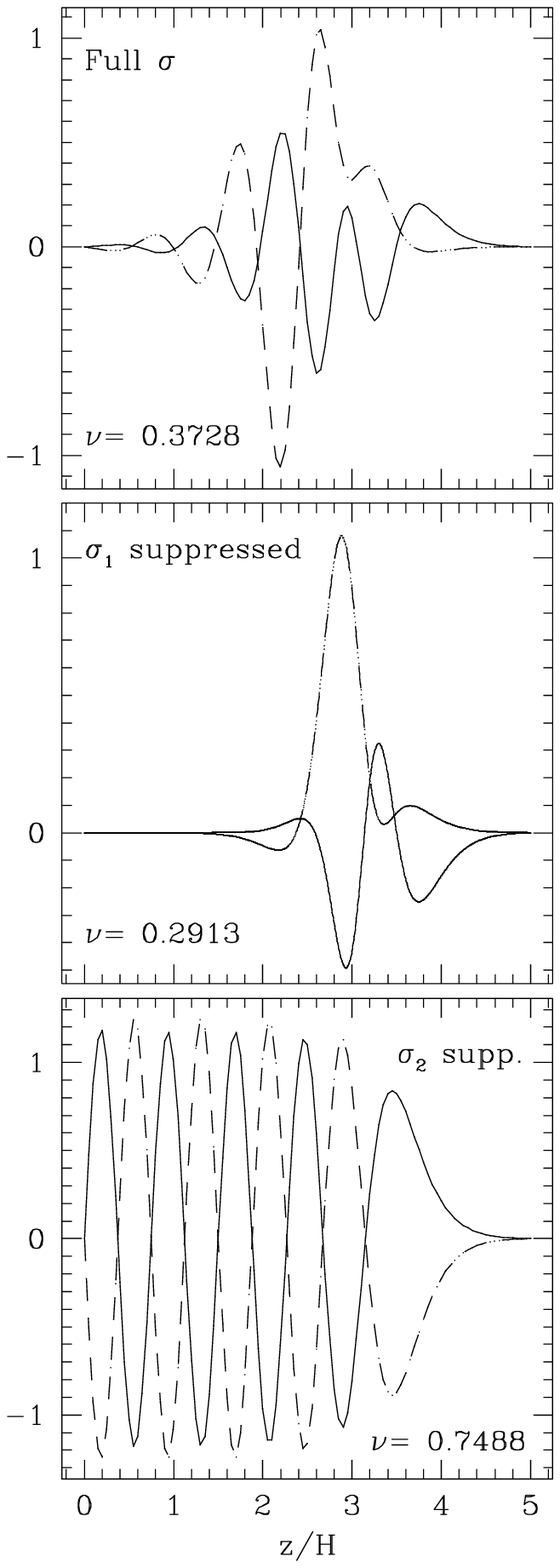}}\vskip 0cm \caption
{Comparison of the structure and growth rate of the MRI for different
configurations of the conductivity tensor for $v_A/c_s = 0.01$.  Top
panel shows the case where both ambipolar diffusion and Hall
($\sigma_1 B_z>0$) conductivity terms are important.  In this
configuration the Hall regime is dominant close to the midplane and
ambipolar diffusion dominates near the surface.  Middle and bottom
panels show the instability under the ambipolar diffusion and Hall
approximations, respectively.  In the top panel $\chi_o=\sqrt 2*0.01$
while in the middle and bottom panels $\chi_o=0.01$.}
\label{fig:real_discs}
\end{figure}

The comparison is presented in Fig.\ \ref{fig:real_discs}.
The structure and growth rate of pure ambipolar diffusion
and Hall ($\sigma_1B_z>0$) perturbations are as expected for $\chi_o \approx
v_A/c_s$, with the ambipolar diffusion decaying towards the
midplane and the envelope of the pure-Hall solutions being fairly
constant.   In both the ambipolar diffusion and the
$\sigma_1=\sigma_2$ cases we obtain a magnetic `dead zone' near the
midplane where no perturbations grow (Gammie 1996, Wardle 1997).
Comparing the top panel of Fig.  \ref{fig:real_discs} with the middle
and bottom ones it is clear that the Hall effect modifies both the
structure and growth of unstable modes.  In particular, the extent of
the dead zone is reduced and the growth rate is increased.  According
to these results, both the depth of the active zones within
the disc and the rate of angular momentum transport by unstable modes
can be significantly modified by the Hall effect.

Although our solutions incorporate the effect of density
stratification on the coupling parameter and the Alfven speed, we have
assumed that the components of the conductivity tensor do not vary
with height.  While this simplification permitted us to compare the
behaviour of the instability in different regimes, the conductivities
in a real disc will reflect the height-dependence of charged particle
abundances and their Hall parameters (see \S \ref{subsec:governing}).
Different regimes are expected to dominate at different heights
(Wardle 2003).

Nonetheless, it is clear from the simplified comparison presented here
that Hall diffusion is an essential part of accretion in
low-conductivity discs, and that it determines the extent of the
magnetically-inactive `dead zone' (Gammie 1996, Wardle 1997).
Further, Hall diffusion will modify any angular momentum transport
within the dead zone that occurs via non-axisymmetric density waves
driven by the active surface layers (Stone \& Fleming 2003) because it
will dominate the marginally magnetically-active regions of the disc
just above the dead zone.  Hall diffusion may therefore affect the
ability of dust grains to settle towards the midplane and begin to
assemble into planetesimals (e.g.\ Weidenschilling \& Cuzzi 1993).

\section{Summary}
\label{sec:summary}

In this paper we have examined the structure and linear growth of the
magnetorotational instability (MRI) in weakly ionised, stratified accretion
discs, assuming an initially vertical magnetic field.  This work is relevant 
for the study of low-conductivity accretion systems, such as protostellar and
quiescent dwarf novae discs, where non-ideal MHD effects are important
(Gammie \&
Menou 1998; Menou 2000; Stone et al. 2000). The formulation allows
for a height-dependent conductivity but in this initial study we assumed 
the components of the conductivity tensor were constant with height.
The analysis was restricted to perturbations with a vertical wavevector
($k = k_z$), which are the most unstable modes when initiated
from a vertically aligned magnetic field (Balbus \& Hawley 1991; Sano \&
Miyama 1999). In this case, the field-parallel component of the conductivity
tensor plays no role and the ambipolar diffusion and resistive limits are
identical.
The linearised system of ODE was integrated from the midplane to
the surface of the disc under appropriate boundary conditions and global
unstable
modes were obtained. The parameters that control the evolution of the fluid
are: (\emph{i}) The coupling between ionised and neutral
components of the fluid evaluated at the midplane ($\chi_o$), which relates
the frequency at which
non-ideal effects are important with the dynamical (keplerian) frequency of
the disc; (\emph{ii}) the magnetic field strength characterised by the ratio
$v_A/c_s$ at the midplane;
and (\emph{iii}) the ratio of the components of the conductivity tensor
perpendicular to the magnetic field $\sigma_1/ \sigma_2$.

In order to explore the growth and structure of unstable modes when 
different conductivity regimes dominate over the entire cross-section of the 
disc, we examined the following configurations of the conductivity tensor:
$\sigma_1 = 0$ (the ambipolar diffusion or resistive limits), $\sigma_2 = 0$
(both Hall limits $\sigma_1 B_z>0$ and $\sigma_1 B_z<0$), and the cases
where both effects are important ($\sigma_1 = \pm \sigma_2$).

The main results of this study are highlighted below:

\begin{enumerate}
\item Global modes are a discrete subset from the continuous curve of
possible $\nu$ vs. $k$ combinations obtained with a local analysis (W99).
These unstable modes can be expressed as a superposition of two WKB
modes with similar growth rate, which explains the interference patterns found
in some of the perturbations.
\item Ambipolar diffusion perturbations peak consistently higher above the 
midplane than solutions where Hall conductivity dominates. 
\item For good coupling ($\chi_o>v_A/c_s$), the structure and growth
of the perturbations are mainly determined by ambipolar diffusion. 
For a weaker coupling, Hall conductivity significantly modifies unstable 
modes. In this case, $\sigma_1=\sigma_2$ perturbations resemble the Hall 
$\sigma_1 B_z>0$ limit and peak closer to the midplane while 
$\sigma_1=-\sigma_2$ modes have their maximum amplitude closer to the surface.
\item Hall limit ($\sigma_1 B_z<0$) perturbations  can have a complex
structure (high wavenumber) even for poor coupling ($\chi_o = 2$). There
are also many unstable modes, which supports findings that in this case the
MRI evolves into MHD turbulence with non-emergence of the two-channel flow
obtained in other regimes (Sano \& Stone 2002a, 2002b).
\item As the coupling parameter $\chi_o$ is reduced, departure from ideal
growth $\nu \sim 0.75$ occurs at a rate that depends on the
conductivity
regime. Hall limit perturbations grow at close to the ideal limit for
$\chi_o>{v_A}^2/{c_s}^2$. In the ambipolar diffusion approximation the growth
rate decreases when $\chi_o \lesssim v_A/c_s$. These results
are in agreement with predictions from W99.
\item The weaker the magnetic field the higher the perturbations peak in all
regimes where ambipolar diffusion is present. On the contrary, both Hall
limits peak closer to the surface with weaker $v_A/c_s$.
\item Unstable modes grow when $v_A/c_s$ is increased
until a critical value ${v_A/c_s}_{crit}$ is reached. At the critical
$v_A/c_s$ the growth rate abruptly drops to zero. At good coupling
${v_A/c_s}_{crit} \sim 1$ for all conductivity regimes. At the poor coupling
limit ($\chi_o = 2$), results are different only for the
Hall regime ($\sigma_1 B_z < 0$). In this case we obtain unstable modes for
$v_A/c_s \sim 2.9$.
\item At very weak magnetic fields ($v_A/c_s \lesssim 0.01$), global effects
are less important, due to the high wavenumber of the perturbations. In this
region of parameter space the growth rates of MRI perturbations tend to the
corresponding local values for the relevant fluid parameters.
\item Hall difussion determines the growth of the MRI when 
$\chi \la |\sigma_1|/\sigma_\perp$. This condition is satisfied over a large 
range of radii in protostellar discs.
\item When the Hall regime dominates near the midplane and ambipolar
diffusion is dominant closer to the surface, a larger section of the disc is
unstable to MRI perturbations and unstable modes grow faster than those
obtained using the ambipolar diffusion approximation
\end{enumerate}

\section{ACKNOWLEDGMENTS}
\label{sec:Acknowledgments}

We thank the referee for useful comments which improved the clarity
of the final paper.

\bsp
\label{lastpage}
\end{document}